\documentclass[prb,twocolumn,showpacs,superscriptaddress]{revtex4-1}
\usepackage[english]{babel}
\usepackage[dvips]{graphicx}
\usepackage{color}
\usepackage{latexsym}
\usepackage{amsmath}
\usepackage{amsthm}
\usepackage{amsfonts}
\usepackage{amssymb}
\usepackage{romannum}

\begin{document}

\pagenumbering{arabic}

\title{Charge-vibration interaction effects in normal-superconductor quantum dots}
\author{P. Stadler}
\affiliation{Fachbereich Physik, Universit{\"a}t Konstanz, D-78457 Konstanz, Germany}
\author{W. Belzig}
\affiliation{Fachbereich Physik, Universit{\"a}t Konstanz, D-78457 Konstanz, Germany}
\author{G. Rastelli}
\affiliation{Zukunftskolleg \mbox{\&} Fachbereich Physik, Universit{\"a}t Konstanz, D-78457, Konstanz, Germany}
\begin{abstract}
We study the quantum transport and the nonequilibrium vibrational states of 
a quantum dot embedded between a normal-conducting and a superconducting lead 
with the charge on the quantum dot linearly coupled to a harmonic oscillator of frequency $\omega$. 
To the leading order in the charge-vibration interaction, 
we calculate the current  and the nonequilibrium phonon occupation 
by the Keldsyh Green's function technique. 
We analyze the inelastic, vibration-assisted tunneling processes in the 
regime $\omega < \Delta$, with the superconducting energy gap $\Delta$, 
and for sharp resonant transmission through the  dot.
When the energy $\varepsilon_0$ of the dot's level is close to the Fermi energy $\mu$, i.e. 
$|\varepsilon_0-\mu| \ll \Delta$,  inelastic vibration-assisted Andreev reflections dominate
up to voltage $eV \gtrsim \Delta$.
The inelastic quasiparticle tunneling becomes the leading process when
the dot's level is close to the superconducting gap $|\varepsilon_0-\mu| \sim \Delta \pm \omega$. 
%
In both cases, the inelastic tunneling processes appear as sharp and prominent 
peaks - not broadened by temperature -  in the current-voltage characteristic 
and pave the way for inelastic spectroscopy of vibrational modes even at temperatures $T \gg \omega$. 
We also found that inelastic vibration-assisted Andreev reflections as well as quasiparticle tunneling 
induce a strong nonequilibrium state of the oscillator. 
In different ranges on the dot's level,  we found that the current produces: 
(i) ground-state cooling of the oscillator with phonon occupation $n \ll 1$, 
(ii)  accumulation of energy in the oscillator with $n \gg 1$ 
and (iii)  a mechanical instability which is a precursor of self-sustained oscillations.  
We show that ground-state cooling is achieved simultaneously for several modes of different frequencies.
Finally, we discuss how the nonequilibrium vibrational state can be readily detected by the asymmetric
behavior of the inelastic current peaks respect to the gate voltage. 
\end{abstract}
%
%
%
%
%
%
%
%
%
%
%
%
%
%
%
%
%
%
%
\date{\today}
\maketitle
%
%
%

%
%
%
%
\section{Introduction}
\label{sec:intro}

Electronic transport through nanoscale devices is characterized by a variety of physical 
phenomena [\onlinecite{Nazarov:2009},\onlinecite{Cuevas-Scheer:2010}].
One of them is the interplay between quantum transport at the level of single electron and the mechanical motion of localized vibrations in various nanoscale devices.
These include single-molecule 
junctions [\onlinecite{Galperin:2007iea,Aradhya:2013et,Park:2000,Yu:2004,Qiu:2004,Wu:2004,Pasupathy:2005,Thijssen:2006,Bohler:2007,deLeon:2008gs,Franke:2012hd}], 
suspended carbon nanotube quantum 
dots (CNT-QDs)[\onlinecite{Nygard:2001cy,Braine:2004hw,Sapmaz:2005fr,LeRoy:2005ke,Sapmaz:2006kv,Leturcq:2009bra,Huttel:2009it,Steele:2009koa,Lassagne:2009fga,Utko:2010eg,Island:2011jq,Ganzhorn:2012kh,Schmid:2012bs,Meerwaldt:2012eb,Island:2012is,Benyamini:2014eb,Weber:2015fa,Deng:2016iw}]
and several types of nanoelectromechanical systems (NEMS) [\onlinecite{Blencowe:2005bz,Greenberg:2012gi,Poot:2012fh}]
as single-electron transistors [\onlinecite{Knobel:2003,Armour:2004ts,Blanter:2004us,Chtchelkatchev:2004cy,Doiron:2006cu}], 
superconducting single-electron
transistors [\onlinecite{LaHaye:2004gs,Naik:2006gf,Clerk:2005ei,Blencowe:2005ih,Koerting:2009jl}], 
single-electron 
shuttles [\onlinecite{Erbe:2001ev,Gorelik:2001hd,Novotny:2003kr,Fedorets:2003bm,Gorelik:2006ej,Smirnov:2004bj,Fedorets:2004hu,Pistolesi:2004ti,Isacsson:2007fk}],  
shuttling nanopilars [\onlinecite{Kim:2013hg},\onlinecite{Prada:2014cd}] and 
quantum dots in suspended 
nanostructures [\onlinecite{Weig:2004eya},\onlinecite{Okazaki:2016bh}]. 
In such nanodevices the motion of the resonator can be not only detected but also manipulated via 
electron transport. 
At the same time, the mechanical motion strongly influences the transport itself.

For instance, in quantum dots, the interaction between the electrons and the harmonic vibrational 
modes gives rise to inelastic tunneling processes appearing 
when consecutive vibrational levels enter the bias window.  
Such an inelastic spectroscopy has been realized in experiments in single molecules [\onlinecite{Yu:2004,Wu:2004,Thijssen:2006,Bohler:2007,deLeon:2008gs}]  
and in suspended 
CNT-QDs [\onlinecite{Sapmaz:2005fr,LeRoy:2005ke,Sapmaz:2006kv,Leturcq:2009bra,Huttel:2009it}]
at least for the high-frequency vibrational modes $\omega \gg T$ ($\hbar=k_B=1$), viz. longitudinal or radial modes, 
but not for the low-frequency flexural modes.
As increasing the bias voltage, tunneling electrons can affect strongly 
the harmonic vibration leading to current-induced nonequilibrium vibrational states,  
in some cases up to the threshold of molecular dissociation [\onlinecite{Franke:2012hd}].

%
%
%
\begin{figure}[b]			
  	\includegraphics[width=0.65\linewidth,angle=0.]{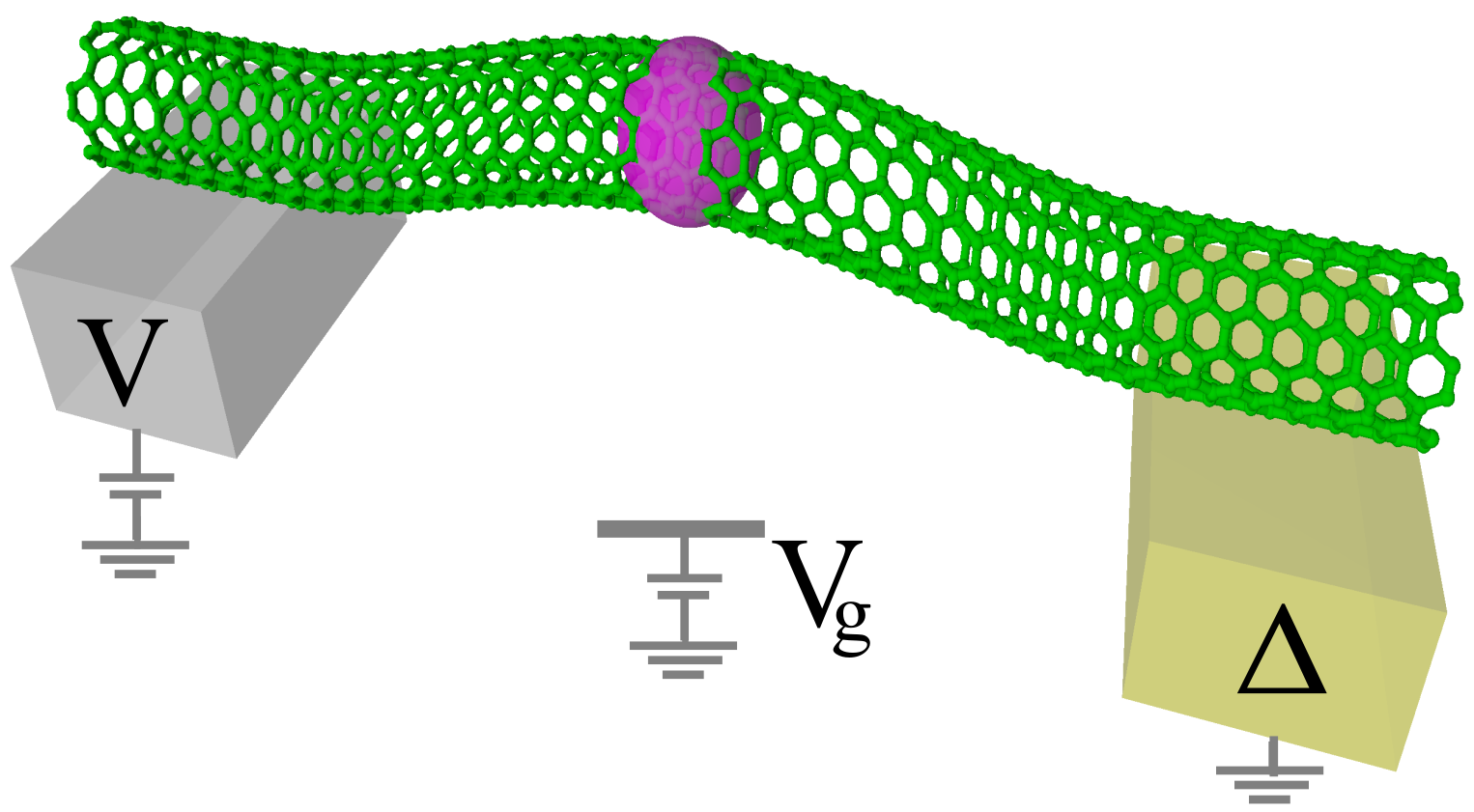} 
  	\caption{Suspended carbon nanotube quantum dot with 
	 the dot's charge coupled to the vibrational flexural modes. The nanotube is in contact with a normal conductor at voltage V 
	 and a superconductor of gap $\Delta$. The dot's energy level is controlled by the gate voltage $V_g$.
	 }  
  	\label{fig:NQDS}
\end{figure}
%
%
%

More generally, charge-vibration interaction leads to a plethora of novel and unexpected behaviors of the nanoresonators.
Electromechanical backaction effects - as oscillator's frequency shift    
and electromechanical damping - have been reported in experiments for the flexural modes in  
suspended CNT-QDs [\onlinecite{Steele:2009koa},\onlinecite{Lassagne:2009fga},\onlinecite{Meerwaldt:2012eb},\onlinecite{Benyamini:2014eb}],  
in quantum dots coupled to a piezoelectric nanoresonator [\onlinecite{Okazaki:2016bh}]
and in a dc-SQUID integrating a flexural resonator [\onlinecite{Poot:2010bc}]. 
Further increasing the coupling strength, current suppression is expected 
due to the Franck-Condon blockade mechanism which was  experimentally observed in CNT-QDs [\onlinecite{Leturcq:2009bra}].
Mechanical bistabilities and  blocked-current states are also theoretically predicted beyond a critical threshold 
of the charge-vibration coupling strength for low frequency (classical) 
nanoresonators [\onlinecite{Galperin:2005ic,Mozyrsky:2006vd,Pistolesi:2007kt,Pistolesi:2008js,Micchi:2015hq,Micchi:2016ft}].  
Other theoretical studies reported negative electromechanical damping [\onlinecite{Usmani:2007bu}], 
self-sustained mechanical oscillations [\onlinecite{Sonne:2008wm,Radic:2011iea,Atalaya:2012,VikstroemPRL2016a}] and charge-induced non-linear effects [\onlinecite{Meerwaldt:2012eb},\onlinecite{Nocera:2012bu},\onlinecite{Nocera:2013gt}].

The majority of experimentally realized single-molecule junctions - as well as suspended CNT-QDs - 
 can be properly modeled  
as single quantum dots [\onlinecite{Galperin:2007iea}].
Hence, the single-impurity 
Holstein model [\onlinecite{Glazman:1988,Wingreen:1989,KoenigPRL1996a,Boese:2001,Lundin:2002bt,Kuo:2002gz,McCarthy:2003cz,Braig:2003cb,Flensberg:2003je,Galperin:2004bs,Mitra:2004ema,Wegewijs:2005hi,Kaat:2005ed}]
has become the paradigmatic model to discuss the effects 
of charge-vibration interaction in such systems.
In its simple formulation, one assumes  a linear coupling between the electron occupation on the dot  
and the oscillation's amplitude of one or more harmonic oscillators representing the local vibrations.
This model has been theoretically investigated widely in literature with different variations,  extensions, 
in different regimes and with various theoretical 
approaches [\onlinecite{Koch:2005hk,Koch:2005if,Zazunov:2006ex,Koch:2006vq,Koch:2006hu,Hwang:2007ie,Luffe:2008wu,Tahir:2008jj,EntinWohlman:2009iza,Haupt:2009cua,Haupt:2010iza,Schultz:2010eb,Cavaliere:2010ks,Yar:2011wv,Piovano:2011dt,Novotny:2011bsa,Fang:2011ek,Li:2012ik,Skorobagatko:2012hx,Rastelli:2012dh,Santamore:2013hg,Hartle:2013dt,Knap:2013dc,Walter:2013cp,Agarwalla:2015dt,Schinabeck:2016cv,Sowa:2017ez}], 
in particular using diagrammatic 
techniques [\onlinecite{Ryndyk:2006ih,Zazunov:2007hva,Egger:2008hha,Dash:2010cz,Rastelli:2010ioa,Dash:2011ez,Souto:2014cm,Chen:2016he}]
and numerical and non-perturbative methods [\onlinecite{Muhlbacher:2008kz,Hutzen:2012dn,Eidelstein:2013io,Jovchev:2013ki,Souto:2015ho}].

However, all previous mentioned theoretical studies of the Holstein model focused on the case of a quantum dot in contact with two normal-conducting leads (N-QD-N). 
Other theoretical works investigated the problem of quantum dots  coupled to the local vibrations and between 
two superconductors (S-QD-S) as a Josephson dot coupled to several bosonic modes [\onlinecite{Novotny:2005kk}],  
a suspended carbon-nanotube acting as nanomechanical 
resonator [\onlinecite{Sonne:2008wm,Zazunov:2006go,Zazunov:2006fz,Sonne:2010tr,Sonne:2010hz,Sadovskyy:2010kf,Padurariu:2012jb}] 
or a molecular Josephson junction model  [\onlinecite{Wu:2012eb}]. The latter was used to explain the resonance peaks 
experimentally observed in vibrating Nb nanowires [\onlinecite{Marchenkov:2007jm},\onlinecite{Kretinin:2013}].

On the contrary,  the hybrid case of a normal-dot-superconductor (N-QD-S)  coupled to the 
bosonic modes  - as  photons [\onlinecite{Sun:1999ve},\onlinecite{Baranski:2015ik}] or 
phonons [\onlinecite{Zhang:2012ie,Wang:2013kca,Dong:2017ce}]  - 
has been less studied in the literature and is theoretically unexplored for the nonequilibrium regime 
of the oscillator.
In this system, one expects that the role of the superconducting lead is unessential for  $\Delta \ll \omega$ 
with the superconducting gap $\Delta$, as the vibrational frequency $\omega$
 sets the energy scale of the inelastic tunneling of electrons. 
The interesting and relevant regime is thus expected for $\Delta \sim \omega$ or $\Delta > \omega$, which corresponds to 
an unrealizable condition  in single-molecule junctions characterized by high-frequency vibrations.
This regime can occur readily and naturally in suspended CNT-QDs, see Fig.~\ref{fig:NQDS}. 
Moreover, CNT-QDs can be inserted between different types of nanocontacts, 
as a normal metal and a superconductor [\onlinecite{Pillet:2013hn,Schindele:2014fm,Gramich:2015dk,Gramich:2016bs,Gramich:2016arxiv}].

In a previous recent work [\onlinecite{Stadler:2016gl}] we explored the N-QD-S system  with charge-vibration interaction  
in the deep subgap regime corresponding to a gap much larger than all 
energy scales (viz. $ \Delta \rightarrow \infty$).  
Our main findings was that the vibrational modes can be cooled to the ground state by inelastic vibration-assisted Andreev reflections  
with phonon number occupation $n\ll 1$  up to the backaction limit dictated by the shot noise and given 
by $n_{BA} \sim {\left( \Gamma_N / \omega \right)}^2$  with the tunneling rate $\Gamma_N$ to the normal lead. 
We also showed that such cooling is exploitable even for several mechanical modes. 
Ground-state cooling by electron transport in suspended CNT-QDs appears as a promising 
strategy  to achieve the quantum regime of low-frequency flexural modes $\omega \ll T$ 
and it has been the subject of a notable research activity during the past 
years [\onlinecite{Lundin:2004cq,Pistolesi:2009ks,Zippilli:2009gp,Zippilli:2010cm,Li:2011kn,Santandrea:2011gh,Arrachea:2014cm,Stadler:2014hra,Stadler:2015dga,Stadler:2016gl}].
We recall that cooling by electron transport has been experimentally demonstrated in 
nanomechanical beams integrating normal metal-insulator-superconductor 
tunnel junctions [\onlinecite{Koppinen:2009bn},\onlinecite{Pekola:2007kn}].

%
%
%
%
\begin{figure}[t!]			
  	\begin{minipage}{\columnwidth}
  		\includegraphics[width=0.85\linewidth,angle=0.]{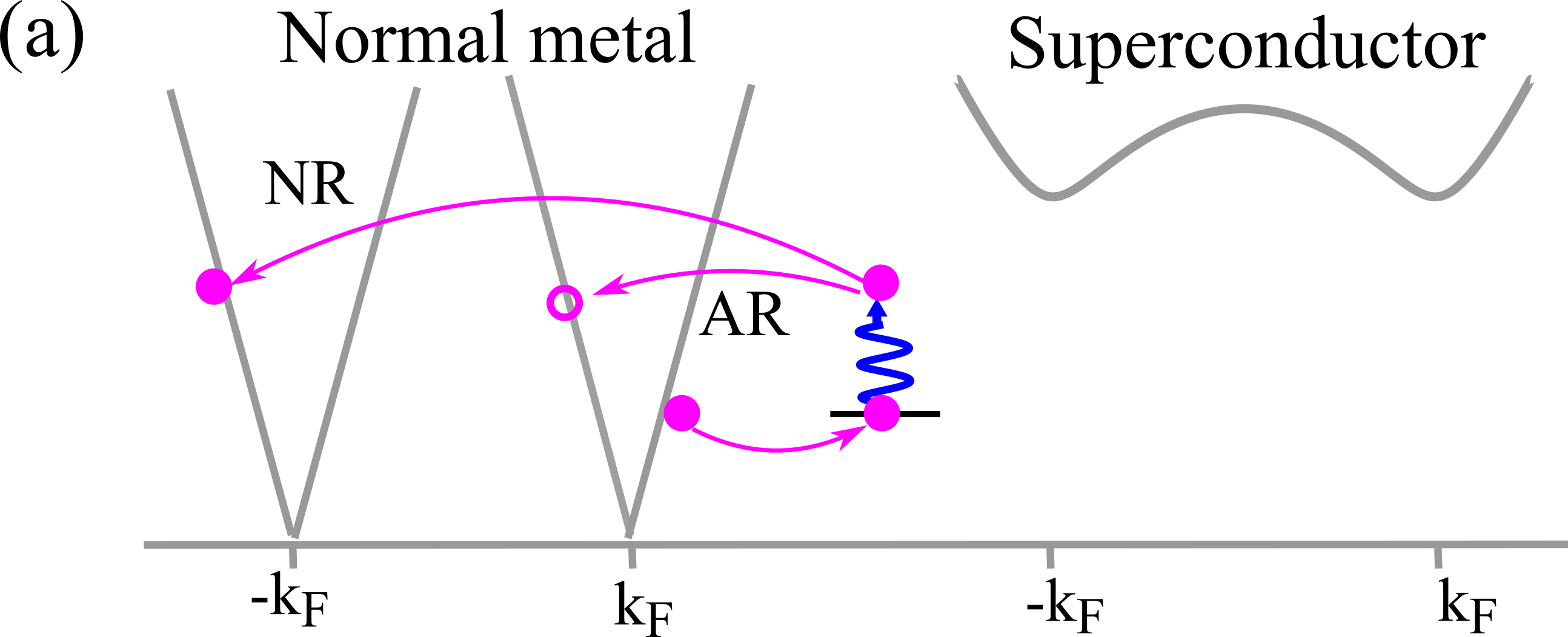} 
  		\vspace{0.5cm}
  	\end{minipage}
  	\begin{minipage}{\columnwidth}
  		\includegraphics[width=0.85\linewidth,angle=0.]{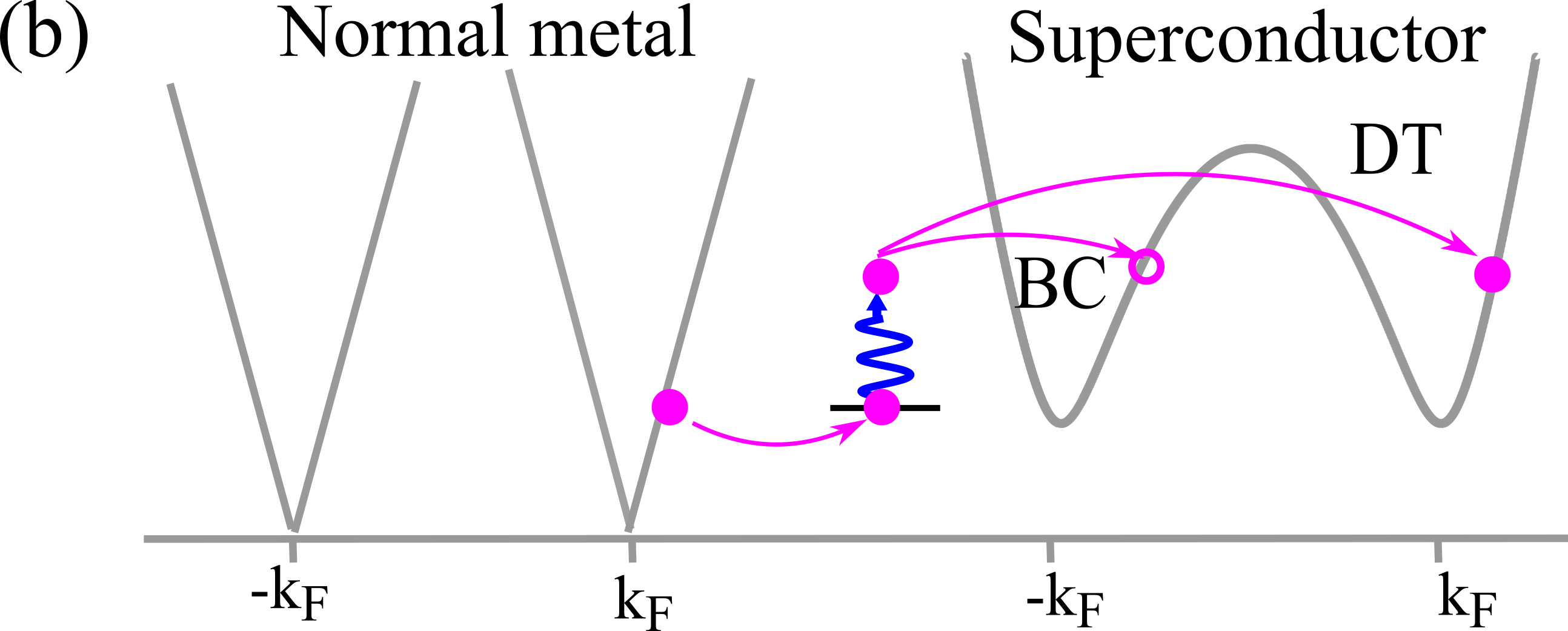}
  	\end{minipage}
  	\caption{Examples of inelastic vibration-assisted tunneling processes 
	occurring in a N-QD-S system with charge-vibration interaction. 
	The electron(hole)-like excitation are sketched as filled (empty) circles respectively. 
An electron-like excitation incoming from the normal metal tunnels to the quantum dot and absorbs a phonon with frequency $\omega$ (wiggled vertical arrow). (a) Below the gap an incoming quasiparticle on the normal metal can be either Andreev reflected (AR) or normal reflected (NR). 
(b) Above the gap, in addition to AR and NR, the incoming quasiparticle can tunnel through the junction by direct tunneling (DT) or branch crossing (BC).
  	}  
  	\label{fig:inelastic_BTK}
\end{figure}
%
%
%
%
%

The objective  of this work is to extend the calculations of Ref.~[\onlinecite{Stadler:2016gl}] 
and consider a finite superconducting gap $\Delta$. 
We aim at demonstrating that the N-QD-S is indeed of particular interest as: 
(i) it is characterized  by a rich and interesting behavior of the inelastic transport   
ruled by inelastic Andreev reflections as well as by inelastic quasiparticle tunneling [see Fig.~\ref{fig:inelastic_BTK}], 
(ii)  it opens the route to inelastic spectroscopy of the low-frequency modes as the flexural ones 
which fundamentally can not be resolved  using normal leads at temperature  $T\gg\omega$,
(iii)  it allows strong manipulation of the vibrational states via electron transport ranging from 
ground-state cooling  up to a mechanical instability, which is possibly a precursor  of self-sustained oscillations. 
Finally, such a system offers intrinsically the possibility to read-out the equilibrium or the 
nonequilibrium state of the vibration directly by the observation of the dc-current.

The paper is structured as follows.
In Sec.~\ref{sec:model} we introduce the Holstein model of a N-QD-S  with charge-vibration interaction 
and discuss the range of validity of our model and of the perturbative diagrammatic expansion. 
%
 
In Sec.~\ref{sec:damping} we discuss the results for the electromechanical damping $\gamma$ showing that it is related 
to several microscopic processes, in particular the inelastic Andreev reflection and inelastic quasiparticle tunneling between 
the normal to the superconducting lead which can drive the vibration out of equilibrium [see Fig.~\ref{fig:inelastic_BTK}].

The phonon number occupation is discussed in the Sec.~\ref{sec:occupation}.
In Sec.~\ref{sec:occupation_single_mode} we discuss the phonon occupation for a single vibrational mode.
Remarkably, in the limit $\omega \ll \Delta$, there is a separation of the ranges of the dot's energy level 
where the inelastic Andreev reflections dominates over the inelastic quasiparticles tunneling or vice-versa.
Both kind of processes can yield ground-state cooling of the oscillator with phonon occupation $n \ll 1$, 
an accumulation of energy in the oscillator with $n \gtrsim1$ and 
a mechanical instability $n \gg 1$ signaled by the negativity of the electromechanical damping.
This region, as well as the region of $n \gg 1$, are beyond our perturbative approach and the validity of harmonicity of the resonator. 
In the Sec.~\ref{sec:occupation_multi_mode} we generalize the discussion taking into account many 
vibrational modes showing that ground-state cooling can be achieved simultaneously for several N modes of different frequencies 
$\omega_n$ via inelastic quasiparticle tunneling [\onlinecite{Stadler:2016gl}].

In the Sec.~\ref{sec:current} we show our results for the current. 
We discuss mainly the inelastic correction to the current and show how the nonequilibrium state of 
the vibration can be  properly read out by the asymmetric
behavior of inelastic current peaks respect to the gate voltage. 
Section \ref{sec:conclusions} contains concluding remarks.

%
%
%
%
\section{Model and Approximation}
\label{sec:model}

In this section, we discuss a model of a quantum dot suspended between a normal-conducting and a superconducting contact 
which can be realized e.g. with a suspended CNT-QD [see Fig.~\ref{fig:NQDS}].

We consider the electrons on the quantum dot occupying a spin-degenerate state with energy $\varepsilon_0$ and the annihilation and creation operators $\hat{d}^{\phantom{g}}_{\sigma}$ and $\hat{d}^\dagger_{\sigma}$ with spin $\sigma$ on the quantum dot. 
The occupation number reads 
$\hat{n}_d = \hat{d}_{\uparrow}^{\dagger}\hat{d}_{\uparrow}^{\phantom{g}} + \hat{d}_{\downarrow}^{\dagger}\hat{d}_{\downarrow}^{\phantom{g}}$.
A single harmonic mode of the resonator with frequency $\omega$ has the bosonic annihilation and creation operators 
$\hat{b}^{\phantom{g}}$  and $\hat{b}^{\dagger}$. 
The electrostatic force between the electrons and a gate voltage induces a coupling between 
the charge and the deflection of the resonator with coupling strength $\lambda$.
The full Hamiltonian is given by
\begin{equation}
\hat{H} = 
\hat{H}_N + \hat{H}_S + \hat{H}_T + \varepsilon_0 \hat{n}_d  
+  \lambda \hat{n}_d (  \hat{b}^{\dagger} +\hat{b}^{\phantom{\dagger}}  )
+ \omega  \hat{b}^\dagger \hat{b}^{\phantom{\dagger}}   \, ,
\label{eq:H}
\end{equation}
with the last two terms describing the charge-vibration coupling and the resonator.
The Hamiltonians of the normal and the superconducting leads are given by
$\hat{H}_N = \sum_{k\sigma} \xi_{N,k} \hat{c}^\dagger_{k\sigma}  \hat{c}^{\phantom{\dagger}}_{k\sigma} $
and
$
\hat{H}_{S} = 
\sum_{k\sigma} 
(
\xi_{S,k}^{\phantom{\dagger}} \hat{a}_{k\sigma}^\dagger \hat{a}_{k \sigma}+\Delta\hat{a}_{-k\downarrow}^\dagger\hat{a}_{k\uparrow }^\dagger+\Delta \hat{a}_{k\uparrow}\hat{a}_{-k\downarrow}
)
$
with the real pairing potential $\Delta$ (superconducting gap) 
and the energy referring to the chemical potential $\xi_{\alpha,k}=\varepsilon_k- \mu_{\alpha}$
with $\alpha=(N,S)$. 
The annihilation operators with spin $\sigma$ and momentum $k$ are given 
by  $\hat{a}_{k\sigma}$ for the superconducting and $\hat{c}_{k\sigma}$ for the normal-conducting lead. 
The tunneling Hamiltonian is 
$
\hat{H}_T = 
\sum_{k\sigma} (
t_{N}^{\phantom{\dagger}} \hat{c}^\dagger_{k\sigma} \hat{d}^{\phantom{\dagger}}_\sigma
+ 
t_{S}^{\phantom{\dagger}} \hat{a}^\dagger_{k\sigma} \hat{d}^{\phantom{\dagger}}_\sigma + \mathrm{H.c.} )
$
with the tunneling energies between the leads and the dot $t_{N}$ and $t_{S}$. 
In the rest of our analysis we assume the wide-band approximation and write 
the tunnel couplings between the quantum dot with the normal contact and with the superconducting contact as 
$\Gamma_N = \pi {\left| t_{N} \right|}^2 \rho_N$ and $\Gamma_S = \pi {\left| t_{S} \right|}^2 \rho_S$, respectively.
The electronic density of states at the Fermi level of the normal lead and of the superconducting lead in the normal (non superconducting) phase are denoted by $\rho_N$ and $\rho_S$.

Before we conclude this section, we discuss the role of the Coulomb interaction in the dot.
In the past, the Anderson model with local Hubbard repulsion $U$ in a N-QD-S system {\sl without charge vibration interaction} 
was studied to understand  
the  competition between superconducting pairing, Coulomb repulsion 
and Kondo correlations [\onlinecite{Fazio:1998dl,Schwab:1999gq,Clerk:2000ge,Cuevas:2001kt,Tanaka:2007ep,MartinRodero:2011hy}].
Analytic methods were restricted to truncated approximated expansions in the regime $U\rightarrow \infty$ 
and a finite superconducting gap  [\onlinecite{Fazio:1998dl,Schwab:1999gq,Clerk:2000ge}]. 
An interpolation approach was used in Ref.~[\onlinecite{Cuevas:2001kt}] for a wide range of the ratio of $U/\Gamma$, 
showing that, for $U/\Gamma < 1$, superconducting pairing dominates the behavior of the quantum dot. 
Tanaka et al. [\onlinecite{Tanaka:2007ep}] employed a numerical renormalization group method in the limit $\Delta\rightarrow \infty$
which allows to map the problem to the case of a QD with a local pairing amplitude $\Delta_D = \Gamma_S$. 
We recovered the same effective Hamiltonian in our previous work [\onlinecite{Stadler:2016gl}], viz. 
$\hat{H}_{dS} = \varepsilon_0 \sum_{\sigma} \hat{d}_{\sigma}^{\dagger } \hat{d}_{\sigma}^{\phantom{g}} - \Gamma_S
(\hat{d}_{\downarrow}^{\phantom{g}} \hat{d}_{\uparrow}^{\phantom{g}} + \mbox{h.c.})$. 
Reference~[\onlinecite{Tanaka:2007ep}]  - limited in the linear transport regime -  points out 
that the effect of the interaction is to renormalize the coupling $\Gamma_S$ 
and the dot level $\varepsilon_0$ in the regime $U \alt \Gamma_S$. 

Tackling the problem of systems with Coulomb interaction and charge-vibration shows a formidable task 
beyond the scope of this work. 
Here we assume the tunneling rate with the normal lead $\Gamma_N$ as 
the lowest energy involved in the problem and the condition for the energy scales $U \alt \Gamma_S < \Delta$  
which sets a conservative estimation for the range of validity of our  approach.
Finally, we remark that experiments with $U \alt \Delta$ as in Refs.~[\onlinecite{Gramich:2015dk},\onlinecite{Gramich:2016bs},\onlinecite{Deacon:2010jn}] 
focused on devices with large asymmetry $\Gamma_S \gg \Gamma_N$ basically 
providing a spectroscopic probe  of the Andreev spectra. The results can be  qualitatively explained in terms of 
a non-interacting model, viz. the Kondo effect is suppressed by the strong proximity effect and correlation effects seems to play no role.

\subsection{Exact results for the quatum dot for $\lambda=0$}
\label{subsec:Gdot}
In this subsection we recall the exact results for the electronic 
Green's function on the dot without charge-vibration interaction and 
the behavior of the density of state of the quantum dot [\onlinecite{Cuevas:1996a},\onlinecite{Blonder:1982a}].  
These Green's function are the building block by which we can express the electromechanical damping, 
the phonon occupation, the zero-order current and the current corrections in presence of the
charge-vibration interaction. 

In Keldysh space, we write the electronic Green's function as
\begin{equation}
\check{G}(t,t^\prime) = 
\begin{pmatrix} \hat{G}^R(t,t^\prime) & \hat{G}^K(t,t^\prime) \\ 0 & \hat{G}^A(t,t^\prime) \end{pmatrix} \, ,
\end{equation}
with the retarded (R), advanced (A) and Keldysh (K) Green's functions defined as
$\hat{G}^{R}(t,t^\prime)  =-i\theta(t-t^\prime) \langle \{\Psi_d(t) , \Psi^\dagger_{d}(t^\prime) \} \rangle$, 
$\hat{G}^{A}(t,t^\prime) =i\theta(t^\prime-t) \langle \{\Psi_d(t) , \Psi^\dagger_{d}(t^\prime) \} \rangle$, 
and
$\hat{G}^{K}(t,t^\prime)  = -i \langle [\Psi_d(t) , \Psi^\dagger_{d}(t^\prime) ] \rangle $. 
Here, we introduced the spinor in electron-hole space as
$
\Psi^\dagger_{d}(t) = (d^\dagger_{\uparrow}(t) \, d_{\downarrow}(t))
$
acting in the Nambu space and denote the commutator 
and anti-commutator with $[\,,\,]$  and $\{\,,\,\}$, respectively.
Notice the notation $\check{}$ for the matrix  defined on the Keldysh contour space and 
$\hat{}$ for the matrix defined on the Nambu space.

The Dyson equation for the Green's function on the dot is
$
\check{G}(\varepsilon)=\check{g}(\varepsilon)+\check{g}(\varepsilon)[\check{\Sigma}_N(\varepsilon)+\check{\Sigma}_S(\varepsilon)]\check{G}(\varepsilon) \, ,
$
with the unperturbed Green's function of the 
dot $\check{g}(\varepsilon)$ and the self-energies $\check{\Sigma}_\alpha(\varepsilon)$ of the normal and superconducting lead [$\alpha=(N,S)$].

In the following, we apply the voltage on the normal-conducting lead $\xi_{N,k} = \varepsilon_k  - \mu -eV$,
$ \xi_{S,k} = \varepsilon_k  - \mu$ and let $\mu=0$.
Using the wide-band approximation, the retarded and Keldysh self-energies 
of the normal metal are $\hat{\Sigma}^R_N=-i\Gamma_N$ and $\hat{\Sigma}_N^K(\varepsilon)=-2i \Gamma_N [1-2\hat{F}(\varepsilon)]$
with the matrix  function 
\begin{equation}
\hat{F}(\varepsilon)=
\begin{pmatrix} f_{+}(\varepsilon) & 0 \\ 
0 & f_{-}(\varepsilon) 
\end{pmatrix} \, ,
\end{equation}
in which we define the two functions $f_{\nu}(\varepsilon)=\{1+\mathrm{exp}[(\varepsilon-\nu\,eV)/T]\}^{-1}$ with the index $\nu=\pm=(e,h)$ to account for the sign of the charge for the electrons and holes, respectively.
The retarded and Keldysh self-energies of the superconductor are
\begin{equation}
\hat{\Sigma}_S^R(\varepsilon)=
-\frac{\Gamma_S}{\Omega(\varepsilon)} 
\begin{pmatrix}
\varepsilon+i\eta & -\Delta \\ -\Delta & \varepsilon+i\eta 
\end{pmatrix}
\end{equation}
and
$
\hat{\Sigma}_S^K(\varepsilon)= [1-2f_0(\varepsilon)] [\hat{\Sigma}_S^R(\varepsilon)-\hat{\Sigma}_S^A(\varepsilon)]
$
with 
$
\Omega(\varepsilon) = \sqrt{\Delta^2-(\varepsilon+i\eta)^2}
$,
an infinitesimal small real part $\eta$ and the Fermi function of the superconducting lead $f_0(\varepsilon)$ $(\nu=0)$.
Solving the Dyson equation, we obtain the retarded Green's function of the dot 
\begin{equation}
\hat{G}^R(\varepsilon) = \left(
\begin{array}{cc}
{G}_{+}(\varepsilon) & {F}(\varepsilon) \\
{F}(\varepsilon) & {G}_{-}(\varepsilon)
\end{array}
\right)
\, , 
\label{eq:GRexact}
\end{equation}
with 
\begin{eqnarray}
G_{\pm}(\varepsilon) 	&=& - \frac{1}{D(\varepsilon)} \left[ \varepsilon\pm \varepsilon_{0}+i \Gamma _N+\varepsilon  \Gamma_S /\Omega(\varepsilon) \right]  	\, , \\
F(\varepsilon) 			&=&    \frac{1}{D(\varepsilon)}  \left[ \Delta  \Gamma_S /\Omega(\varepsilon) \right] 	\, 
\end{eqnarray}
and
\begin{equation}
\hspace{-2mm} D_{}(\varepsilon)
\! = \! 
\frac{\Delta^2 \Gamma_S^2}{\Omega^{2}(\varepsilon)}
\!-\!
\left[ \varepsilon \! +\! \varepsilon_{0} \! + \! i \Gamma_N \! + \! \frac{\varepsilon  \Gamma_S}{\Omega(\varepsilon)} \! \right] 
\left[ \varepsilon  \!- \! \varepsilon_{0} \! + \! i \Gamma_N \! + \! \frac{\varepsilon  \Gamma_S}{\Omega(\varepsilon)} \!  \right]
\label{eq:Dexp}
\end{equation}
The Keldysh Green's function of the dot is obtained from the Dyson equation and is given by 
$
\hat{G}^K(\varepsilon)=\hat{G}^R(\varepsilon)(\hat{\Sigma}_N^K(\varepsilon)+\hat{\Sigma}_S^K(\varepsilon))\hat{G}^A(\varepsilon)
$.

%
%
%
%
\begin{figure}[t!]
  	\includegraphics[width=0.47\columnwidth,angle=0.]{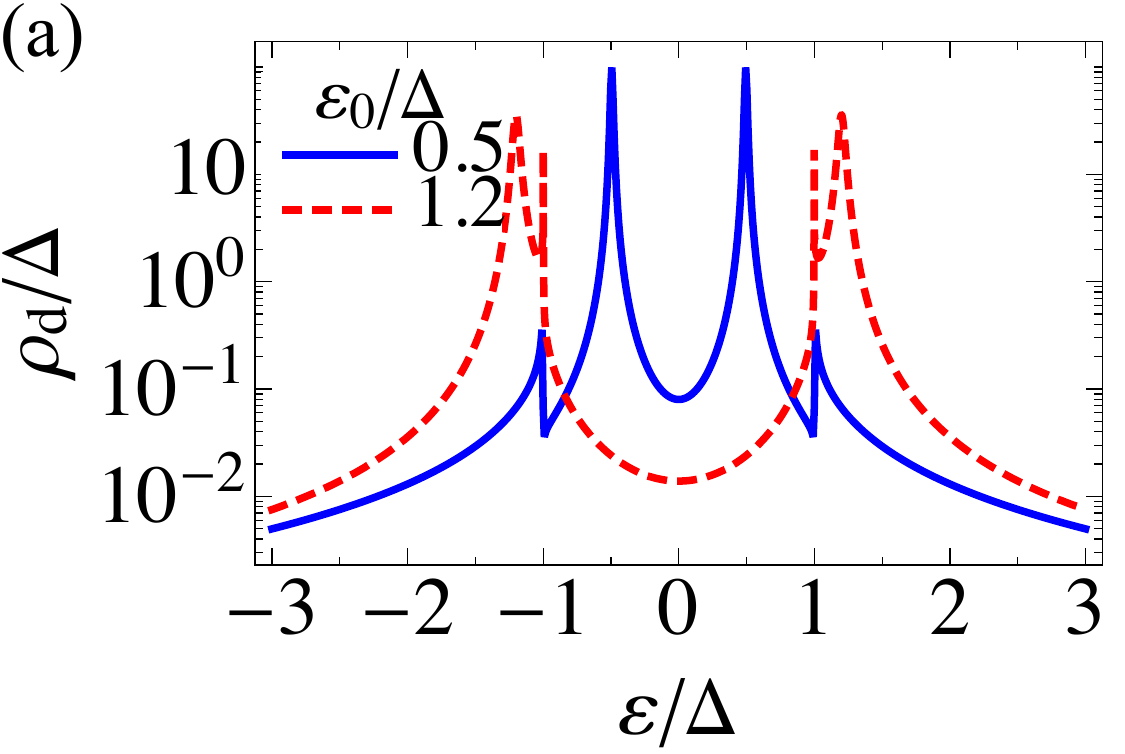} \includegraphics[width=0.47\columnwidth,angle=0.]{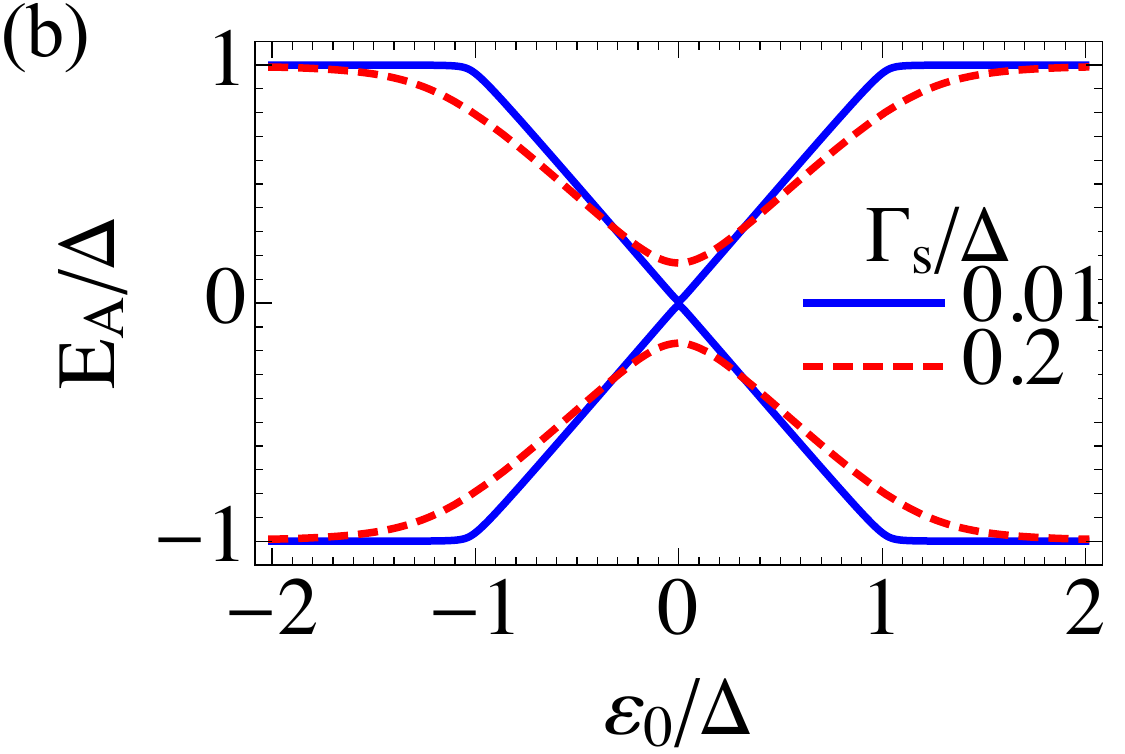} 
  	\caption{(a) Density of states of the quantum dot $\rho_d(\varepsilon)$
	as a function of $\varepsilon$ at $\Gamma_N=\Gamma_S = 0.1 \Delta$ 
	and for different bare dot's level $\varepsilon_0$. 
	(b) Andreev levels $E_A$ as a function of $\varepsilon_0$ for $\Gamma_N=0.01\Delta$ 
	 and different  $\Gamma_S$.
	For $\varepsilon_0>\Delta$, the Andreev levels are very close to the gap edge and, additionally, 
	a resonant peak occurs at $\varepsilon_0$.
	}  
  	\label{fig:DensityOfStates} 
\end{figure}
%
%
%
%
%

The density of states $\rho_d(\varepsilon)$ on the quantum dot is calculated by 
\begin{equation}
\rho_d(\varepsilon) =  -\textrm{Im}\,\textrm{Tr}[\hat{G}^R(\varepsilon)] \, ,
\end{equation}
with the retarded Green's function given by Eq.~\eqref{eq:GRexact}. 
An example of the density of states as a function of energy is shown in Fig.~\ref{fig:DensityOfStates}(a). 
For $|\varepsilon| \gg \Delta$, $\rho_d(\varepsilon)$ shows a similar behavior as for a quantum dot coupled to two normal leads. 
At low energy, the superconducting correlation penetrate on the quantum dot and lead to a suppression of the density of states 
and to the formation of Andreev levels pinned inside the superconducting gap and signaled by the two peaks. 
These Andreev levels $E_A$ have an energy-dependence given by the zeroing of $D(\varepsilon)$ 
in Eq.~\eqref{eq:Dexp} [see  Fig.~\ref{fig:DensityOfStates}(b)]. 
For large superconducting gap, the Andreev levels  are approximated 
by $E_A \simeq \pm \sqrt{\varepsilon_0^2 + \Gamma_S^2}$ 
valid for $|E_A| \ll \Delta$.

\subsection{Phonon Green's function with charge-vibration interaction}

In this section we discuss the phonon Green's function for one harmonic oscillator.
We use the perturbative expansion to calculate the phonon self-energy $\check{\Pi}(\varepsilon)$ to lowest order in the charge-vibration coupling [\onlinecite{Mitra:2004ema}].
In this way, we can obtain the electromechanical damping and, by solving the Dyson equation, the nonequilibrium phonon occupation.
Analytic formulas are possible in the limit of small damping $\gamma \ll \omega$ and a high quality factor of the resonator.
Such an approach is equivalent to solve the quantum Langevin equation for the harmonic oscillator assuming as input the bare quantum noise correlators 
of the dot's charge in the N-QD-S without charge-vibration interaction [\onlinecite{Clerk:2004cz}] and using the 
Markovian approximation.
The latter approximation holds in our case as the typical frequency scale $\delta\omega$ of the frequency depending electromechanical 
damping and of the noise  scales as $\delta\omega \sim \Gamma_N$ and is still larger than then linewidth 
of the oscillator's susceptibility which scales as 
 $\sim \lambda^2 \Gamma_N/\omega_0^2$ in the regime $\Gamma_N < \omega$.

We introduce the symmetrized bosonic operator $\hat{A}(t) = \hat{b}^\dagger(t)+\hat{b}(t)$, the contour ordering operator $T_c$ on the Keldysh contour and define the phonon propagator on the Keldysh contour as $D(\tau,\tau^{\prime}) = -i \langle T_c \hat{A}(\tau) \hat{A}^{\dagger}(\tau^\prime) \rangle$ with the contour variable $\tau$ on the Keldysh contour.  
Then, we transform the contour variable $\tau$ to the real time and use the Larkin-Ovchinnikov rotation 
to write the Dyson equation in the triangular form. The phonon Green's function in Keldysh space are given by
\begin{equation}
\check{D}(t,t^\prime) = 
\begin{pmatrix} D^R(t,t^\prime) & D^K(t,t^\prime) \\ 0 & D^A(t,t^\prime) \end{pmatrix}
\end{equation}
with the phonon Green's functions  
$
D^R(t,t^\prime)  = -i \theta(t-t^\prime) \langle  [ \hat{A}(t),   \hat{A}(t^\prime) ] \rangle
$, 
$
D^A(t,t^\prime)  = i \theta(t^\prime-t) \langle  [ \hat{A}(t),   \hat{A}(t^\prime) ] \rangle
$,  and
$
D^K(t,t^\prime)  = -i  \langle  \{\hat{A}(t^\prime), \hat{A}(t) \} \rangle 
$. 
Finally, we obtain the Dyson equation in Fourier space as
$
\check{D}(\varepsilon) =  \check{D}_0(\varepsilon)  +  \check{D}_0(\varepsilon)\check{\Pi}(\varepsilon) \check{D}_0(\varepsilon).
$
The bare phonon Green's functions in $\check{D}_0(\varepsilon)$ are given by 
$
D^{R,A}_0(\varepsilon)=2 \omega/[(\varepsilon\pm i\eta)^2-\omega^2]
$ 
and
$
D^{K}_0(\varepsilon)=-2\pi i \left[\delta(\varepsilon\mathord-\omega)\mathord+\delta(\varepsilon\mathord+\omega)\right] \mathrm{coth}[\omega/(2T)]
$. 
The polarization $\check{\Pi}(\varepsilon)$ corresponds to the phonon self-energy (polarization diagram) 
to the leading order in the electron-vibration interaction $\sim \lambda^2$
between the resonator and the electrons.  
These self-energies are given by
\begin{eqnarray}
\label{eq:pRA}
\Pi^{R,A}(\varepsilon) &=& \! 
i \frac{\lambda^2}{2}  \!\! \int \!\!  \frac{d\varepsilon^\prime}{2\pi} \mbox{Tr} 
\left[\hat{\tau}_{z} \hat{G}^{K}(\varepsilon^\prime)\hat{\tau}_{z}\hat{G}^{A,R}(\varepsilon^\prime-\varepsilon) \right. \\ 
+ & \hat{\tau}_{z} & \left. \! \hat{G}^{R,A}(\varepsilon^\prime)\hat{\tau}_{z}\hat{G}^K(\varepsilon^\prime-\varepsilon) \right] \, , 
\end{eqnarray}
and
\begin{eqnarray}
\Pi^K(\varepsilon) &=& 
 \!  i \frac{\lambda^2}{2}  \!\! \int \!\! 
\frac{d\varepsilon^\prime}{2\pi} \mbox{Tr} 
\left[\hat{\tau}_{z}\hat{G}^{K}\!(\varepsilon^\prime)\hat{\tau}_{z}\hat{G}^{K}(\varepsilon^\prime\!-\!\varepsilon)  \right. \\
+& \hat{\tau}_{z}& \! \left. \hat{G}^{R}\!(\varepsilon^\prime)\hat{\tau}_{z}\hat{G}^{A}\!(\varepsilon^\prime\!-\!\varepsilon) \!+\!
\hat{\tau}_{z}\!\hat{G}^{A}\!(\varepsilon^\prime)\hat{\tau}_{z}\hat{G}^{R}\!(\varepsilon^\prime\!-\!\varepsilon)\right]
\end{eqnarray}
in which the trace $\mathrm{Tr}$ and the Pauli matrix $\hat{\tau}_z$ acts on the Nambu space.

To take into account the intrinsic damping of the oscillator, we additionally include a self-energy $\check{\Sigma}_0(\varepsilon)$ in Keldysh space. From the phenomenological model [\onlinecite{Stadler:2015dga}], one obtains the expressions
\begin{align}
\mathrm{Im}\,\Sigma^R_0(\varepsilon) &= -\varepsilon/Q \\
\Sigma_0^K(\varepsilon) &=-2i\varepsilon \, \mathrm{coth}(\varepsilon)/Q \, ,
\end{align}
in which the coefficient $Q$ corresponds to the quality factor of the resonator. 

Using the Dyson equation, we obtain the retarded, advanced and Keldysh equations
\begin{align}
{D}^{R,A}(\varepsilon) 
&=
\frac{2\omega}{\varepsilon^2-\omega^2-2\omega[{\Pi}^{R,A}_{}(\varepsilon)+\Sigma_0^R(\varepsilon)]} 
\nonumber
\\
 & 
 \simeq
\frac{1}{\varepsilon-{\omega}\pm i\gamma_{\mathrm{tot}}}-\frac{1}{\varepsilon+{\omega}\pm i\gamma_{\mathrm{tot}}} \, , \label{eq:D_RA} \\
{D}^K(\varepsilon) &={D}^R(\varepsilon) [\Pi^K_{}(\varepsilon)+\Sigma_0^K(\varepsilon)] {D}^A(\varepsilon) 
\nonumber \\
&\simeq i\pi \frac{\Pi^K(\varepsilon)+\Sigma^K_0(\varepsilon)}{\gamma_{\mathrm{tot}}(\varepsilon)}[\delta(\varepsilon-\omega)+\delta(\varepsilon+\omega)] \label{eq:D_K} \, .
\end{align}
with the total damping rate given by
\begin{equation}
\gamma_{\mathrm{tot}}(\omega)=-\textrm{Im}\,[\Pi^R(\omega)+\Sigma^R_0(\omega)] = \gamma+\gamma_0 \, .
\label{eq:damping}
\end{equation} 
which can be separated into an intrinsic damping $\gamma_0=-\mathrm{Im}\, \Sigma_0^R(\omega)= \omega/Q$ and the damping $\gamma=-\mathrm{Im}\, \Pi^R(\omega)$ due to the charge-vibration interaction.
In Eq.~\eqref{eq:D_RA} we expanded the retarded and advanced phonon 
Green's function close to $\varepsilon \simeq \omega$ taking into account the frequency shift of the resonator 
$\Delta\omega=\textrm{Re}\,\Pi^R(\omega)$.

%
%
%
%
 
\section{Results for the electromechanical damping}
\label{sec:damping}
We found that the electromechanical damping $\gamma$ in Eq.~\eqref{eq:damping} 
due to the electron-vibration interaction can be divided into basic reflection and transmission 
processes of charge transport occurring in N-QD-S systems [\onlinecite{Cuevas:1996a},\onlinecite{Blonder:1982a}].
After some algebra, we cast  the electromechanical damping   as
\begin{equation}
\gamma=\gamma_{\mathrm{NN}}+\gamma_{\mathrm{NS}}+\gamma_{\mathrm{SS}}\, ,
\label{eq:gamma_tot}
\end{equation}
in which the damping coefficients $\gamma_{\alpha\beta}$ 
with the labels $\alpha=(\mathrm{N},\mathrm{S})$ and $\beta=(\mathrm{N},\mathrm{S})$ 
are associated to an inelastic process with an absorption or emission 
of one phonon with an incoming particle and an outgoing particle 
involving the leads $\alpha$ and  $\beta$. 
Examples of such processes are given in Fig.~\ref{fig:inelastic_BTK}.

The first term in Eq.~(\ref{eq:gamma_tot})  corresponds to reflections  at the normal lead 
and can be divided into inelastic Andreev reflection (AR) and normal inelastic 
reflection (NR),
\begin{equation}
\gamma_{\mathrm{NN}} = \gamma_{\mathrm{AR}}+\gamma_{\mathrm{NR}} \, .
\end{equation}
An schematic view of inelastic AR and NR with an absorption of a phonon before an AR or NR is shown in 
Fig.~\ref{fig:inelastic_BTK}(a) in which the excitation spectra of the normal and the superconducting 
lead around the Fermi momenta $k_F$ are shown as gray lines. 
Below the superconducting gap, these processes are the only possible processes  as 
an incoming electron can only be Andreev reflected or normal reflected.

\begin{figure}[b]			
 	\begin{minipage}{.5\textwidth}
        \includegraphics[width=0.45\linewidth,angle=0.]{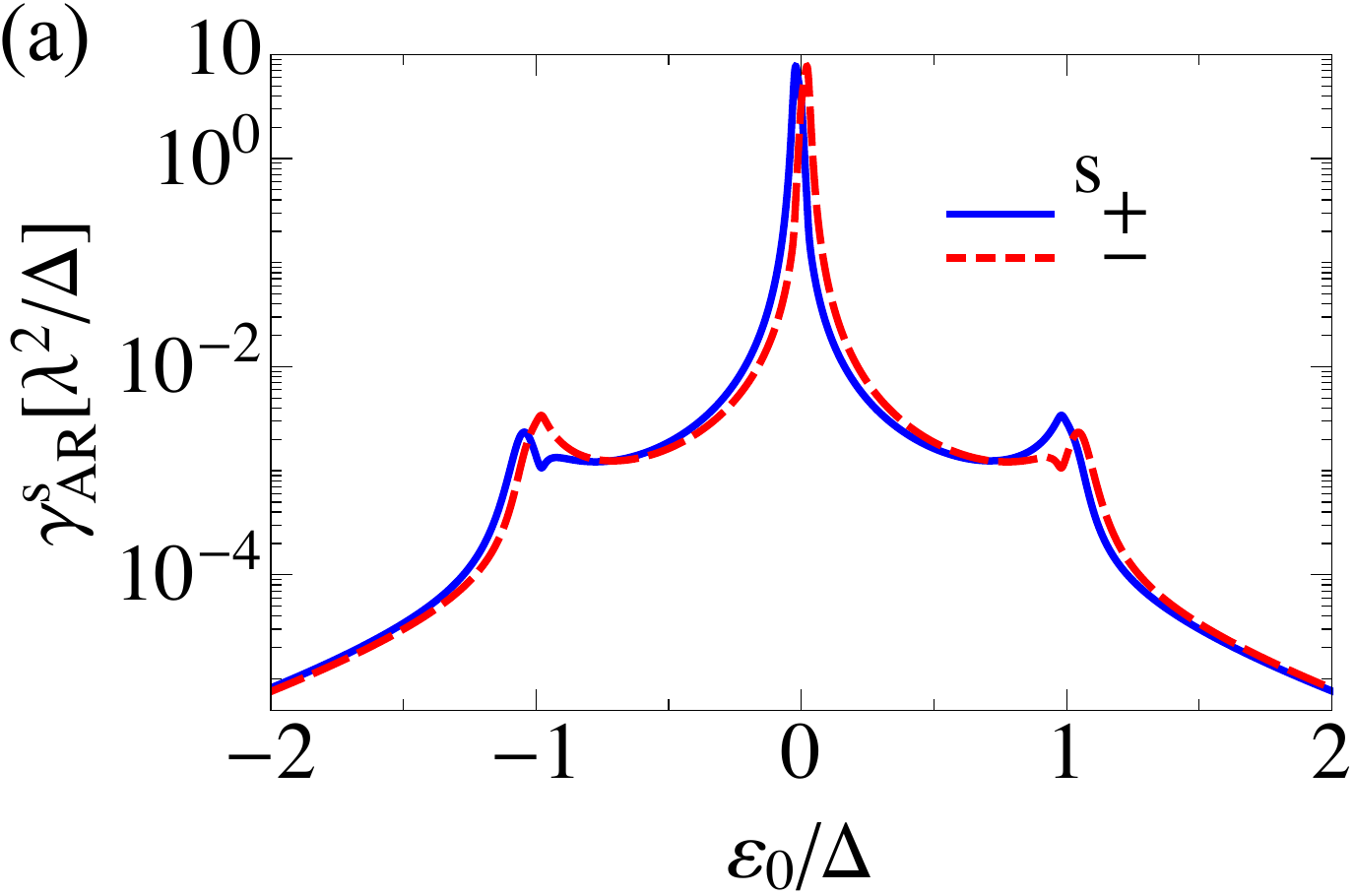}
 		\includegraphics[width=0.45\linewidth,angle=0.]{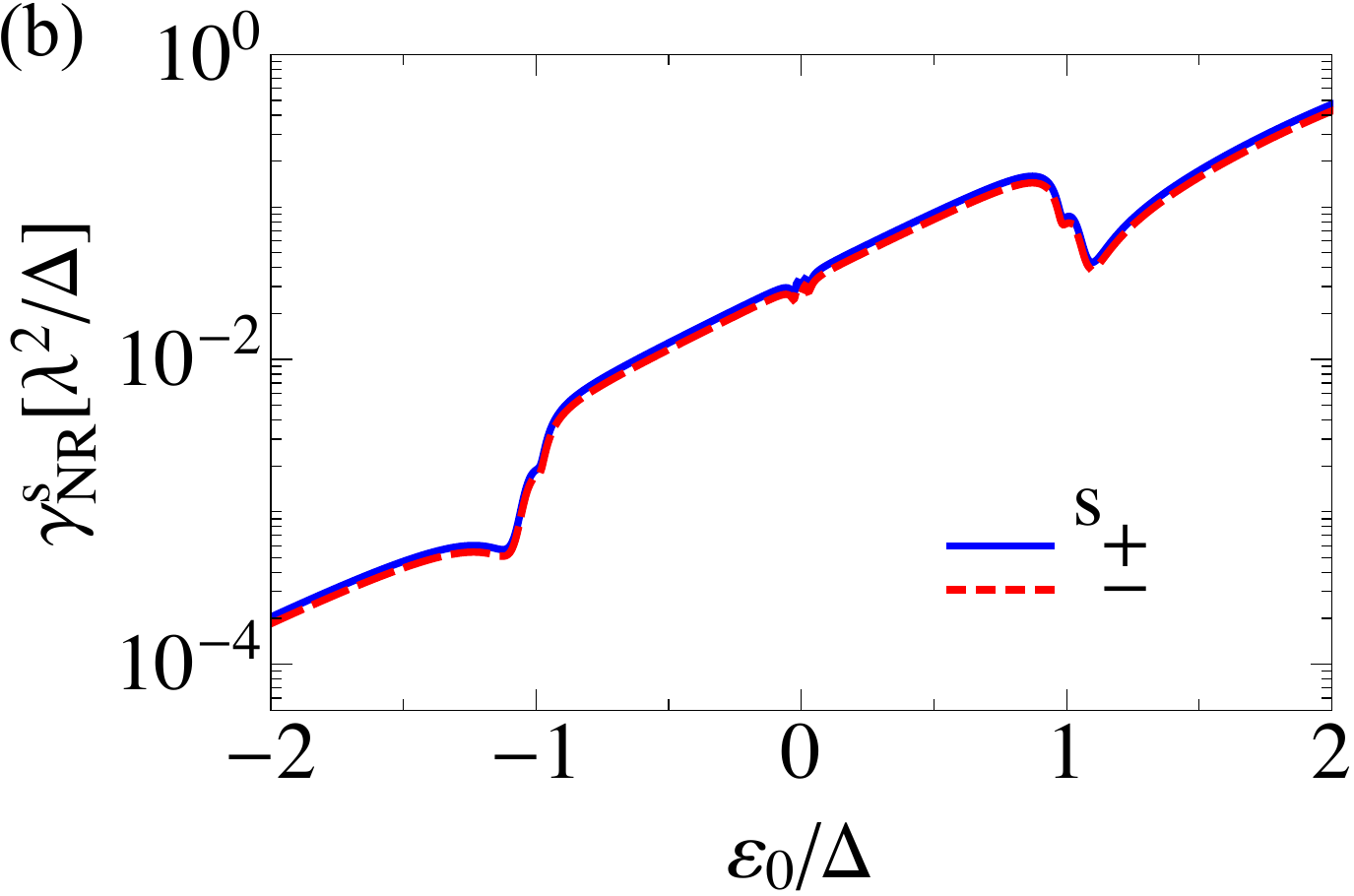} 
	 \end{minipage}	
	 \caption{Electromechanical damping associated to inelastic Andreev reflections  $\gamma_{\mathrm{AR}}^s=\sum_v \gamma_{\mathrm{AR}}^{s\nu}$ and normal reflections $\gamma_{\mathrm{NR}}^s=\sum_v \gamma_{\mathrm{NR}}^{s\nu} $ 
	as a function $\varepsilon_0$ at $eV=3\Delta$. The other parameters are $\Gamma_N=\Gamma_S=0.2\omega$, $\gamma_0=10^{-6}\omega$ and $\omega=0.05\Delta$.
    }  
 	\label{fig:damping_AR_NR}
 \end{figure}

The damping rate  $ \gamma_{\mathrm{AR}}$ and $\gamma_{\mathrm{NR}}$  can be written as the sum 
of elemental processes 
\begin{align}
\gamma_{\mathrm{AR}} &= \sum_{\nu s} s \gamma_{\nu\bar{\nu}}^{s} \, ,   \label{eq:gammaAR} \\
\gamma_{\mathrm{NR}} &= \sum_{\nu s} s  \gamma_{\nu\nu}^{s} \, ,  \label{eq:gammaNR}
\end{align}
with the index $\nu=\pm=(e,h)$ for electrons and holes, respectively, and the notation $\bar{\nu}=-\nu$. 
The index $s=\pm$ refers to an inelastic process with an absorption ($s=+$) and the emission ($s=-$) of one phonon. 
As an example, $\gamma_{eh}^s$ corresponds to an inelastic AR with an absorption or emission of a phonon for an incoming 
electron and an outgoing hole  at the normal lead.
The damping rates $ \gamma_{\nu\nu^\prime}^{s}$ in Eqs.~\eqref{eq:gammaAR} and \eqref{eq:gammaNR} can be written as
\begin{equation}
 \gamma_{\nu\nu'}^{s}(\omega) =   \lambda^2\frac{\Gamma_N^2}{2}
 \int \frac{d\varepsilon}{2\pi}   
 \left| T_{\nu\nu'}^{s}(\varepsilon) \right|^2
 f_{\nu}(\varepsilon)[1-f_{\nu'}(\varepsilon+s\omega)] 
 \, ,
\end{equation}
with the Fermi functions $f_{\nu}(\varepsilon)=1/[1+\mathrm{exp}((\varepsilon-\nu\,eV))/T]$ for electron and holes, respectively, and the transmission functions
\begin{align}
 \label{eq:T_AR}
 T_{\nu\bar{\nu}}^{s}(\varepsilon) \! &= \! 
 {
 	G_{\nu} (\varepsilon) F_{}^{*}(\varepsilon+s\omega)
 	\!\! - \! 
 	F_{}^{}(\varepsilon)   G_{\bar{\nu}}^{*} (\varepsilon+s\omega)  
 }  \, , 
 \\
 \label{eq:T_NR}    
 T_{\nu\nu}^{s}(\varepsilon) \! &= \! 
 { 		G_{\nu} (\varepsilon) G_{\nu}^{*} (\varepsilon+s\omega)
 	\!\! - \! 
 	F_{}^{}(\varepsilon) 
 	F_{}^{*}(\varepsilon+s\omega)  \, .
 } 
\end{align}
As discussed in Ref.~[\onlinecite{Stadler:2016gl}] the transmission functions consist of a coherent sum of two 
amplitudes that are associated to the two possible paths in which the phonon is emitted or 
absorbed before or after a reflection. 
To the lowest order  in the tunneling rates, the electromechanical dampings due to AR and NR are 
proportional to $\sim \Gamma^2_S\Gamma_N^2$  and $\sim  \Gamma^2_N$, respectively. 
An example of the behavior of $\gamma_{\mathrm{AR}}$ and  $\gamma_{\mathrm{NR}}$ is shown in Fig.~\ref{fig:damping_AR_NR}. 
We analyzed in detail these processes in Ref.~[\onlinecite{Stadler:2016gl}].

%
%
%
%
%
%
%
%

We now turn to discuss the term $\gamma_{\mathrm{NS}}$ 
in which a particle is exchanged between the normal contact (electron or hole) 
and the superconducting contact (quasiparticle). 
We can break up these terms in absorption or emission processes with 
a initial electron or hole, namely 
\begin{align}
\gamma_{\mathrm{NS}}&=\sum_{s\nu} s \gamma_{\mathrm{NS}}^{s\nu}
\end{align}
with
\begin{equation}
\gamma_{\mathrm{NS}}^{s\nu}  \!\! =  \!
16 \lambda^2 \Gamma_N \Gamma_S \!\!
\int \!\!\! \frac{d\varepsilon}{2\pi}  
\mathcal{T}^{s}_{\nu}(\varepsilon) 
{\rho}_S(\varepsilon+s\omega) f_{\nu}(\varepsilon)[ \! 1 \! - \! f_0(\varepsilon+s\omega)] \, .
\label{eq:gamma_ns}   
\end{equation}
In Eq.~\eqref{eq:gamma_ns} we have introduced the superconducting density of states defined as
$
{\rho}_S(\varepsilon) = \theta(\varepsilon^2-\Delta^2)\varepsilon/\sqrt{\varepsilon^2-\Delta^2}
$ normalized to the one of the normal phase, and the  transmission function
\begin{equation}
\mathcal{T}^{s}_{\nu}(\varepsilon) 
  \! =  \!
\vert T_{\nu\nu}^{s}(\varepsilon) \vert^2+\vert T_{\nu\bar{\nu}}^{s}(\varepsilon)\vert^2+ \frac{2\Delta}{\varepsilon+s\omega} \textrm{Re}[T^{s}_{\nu\nu}(\varepsilon){T^{s}_{\nu\bar{\nu}}(\varepsilon)}^*] 
\, .
\label{eq:T_s_nu}  
\end{equation}
The processes described by Eq.\eqref{eq:gamma_ns} admit two equivalent descriptions: 
for instance the term for $\nu=e$ can be seen as an impinging electron from the normal lead $N$ to the superconductor $S$, 
leaving an hole in the lead $N$, or it can be seen as an impinging quasiparticle from the superconductor $S$ transmitted as an hole in the normal lead $N$.

%
%
%
%
 \begin{figure}[t]			
		\includegraphics[width=0.7\linewidth,angle=0.]{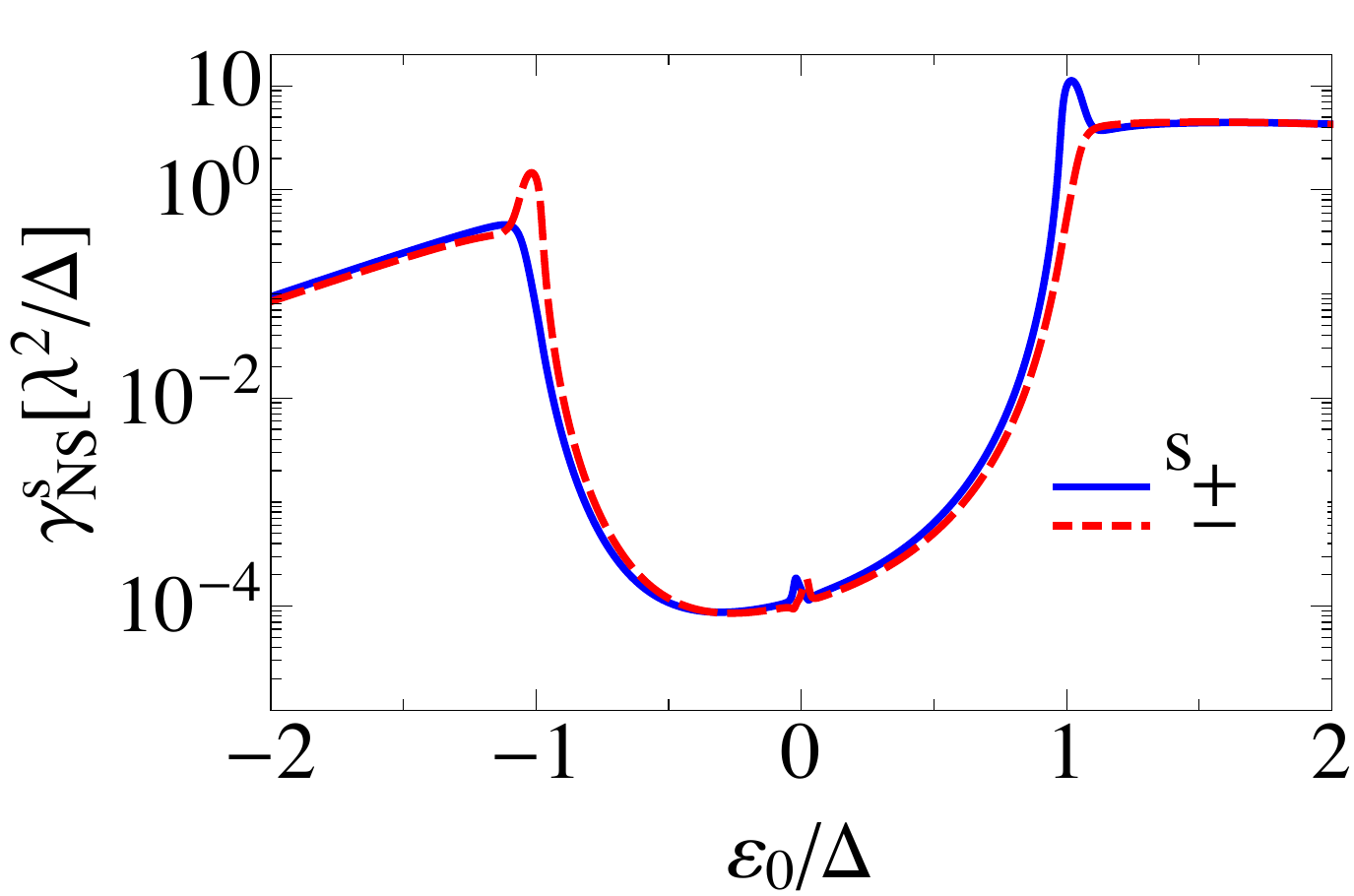} 
 	\caption{Electromechanical damping associated to the inelastic tunneling between N and S $\gamma_{\mathrm{NS}}^{s}=\sum_{\nu}\gamma_{\mathrm{NS}}^{s\nu}$ 
	for one impinging electron $(\nu=+=e)$	
	as a function  $\varepsilon_0$ at $eV=3\Delta$. The parameters are $\Gamma_N=\Gamma_S=0.2\omega$,  
	$\gamma_0=10^{-6}\omega$ and $\omega=0.05\Delta$.
	}  
 	\label{fig:damping}
 \end{figure}
%
%
%
%
%

Comparing the results Eqs.~\eqref{eq:gamma_ns} and \eqref{eq:T_s_nu}  
with the ones reported in literature [\onlinecite{Blonder:1982a},\onlinecite{Cuevas:1996a}], 
it turns out that the term in Eq.~\eqref{eq:T_s_nu}  
proportional to $\vert T_{\nu\nu}(\varepsilon) \vert^2$ corresponds to the 
inelastic direct tunneling of quasiparticles (DT) and the term 
with $\vert T_{\nu\bar{\nu}}(\varepsilon) \vert^2$  corresponds to branch crossing (BC), 
according to the terminology introduced in the BTK model [\onlinecite{Blonder:1982a}].
When an electron is transmitted via inelastic DT, it occupies an electron-like branch on the superconductor. 
Branch crossing refers to the transmission of the electron on the hole-like branch of the superconductor. 
Examples of processes corresponding to inelastic DT and BC for an incoming electron from the normal metal are sketched in Fig.~\ref{fig:inelastic_BTK}(b), in which a phonon is absorbed when the electron is transmitted 
to the superconductor. 
To the leading order in the tunneling rates, DT and BC are proportional to $\Gamma_S \Gamma_N$ 
and $\Gamma_S^2 \Gamma_N^2$, respectively. 
The last term in Eq.~\eqref{eq:T_s_nu} is proportional to the  product of two transmission functions and 
can be interpreted as tunneling via an intermediate state [\onlinecite{Cuevas:1996a}] and scales as $\Gamma_N \Gamma_S^2$.

An example of the behavior of  $\gamma_{\mathrm{NS}}^{s}=\sum_{\nu}\gamma_{\mathrm{NS}}^{s\nu}$ is shown in Fig.~\ref{fig:damping}.
By comparison of Fig.~\ref{fig:damping} with Fig.~\ref{fig:damping_AR_NR} we notice that the damping associated to the inelastic tunneling of quasiparticles is negligible when the dot's level is well inside  
the gap  $|\varepsilon_0| \ll \Delta$ and becomes relevant when $|\varepsilon_0| \sim \Delta$ whereas 
the damping associated to inelastic Andreev reflections dominates at low energy $|\varepsilon_0| \ll \Delta$.
We will explain this behavior in the next section.

Finally, for completeness, we report the expression for the reflections involving the quasiparticles of the superconductor.
This term is given by
\begin{equation}
\gamma_{\mathrm{SS}} = \sum_{\nu s} s \gamma_{\textrm{SS}}^{s\nu}  \, .
\end{equation}
with each term referring to individual process of emission or absorption for electron/hole.
They read   
\begin{equation}
	\gamma_{\textrm{SS}}^{s\nu} \! = \! 
	8 \lambda^2 \Gamma_S^2 \! \int  \! \frac{d\varepsilon}{2\pi} 
	\mathcal{R}^{s}_{\nu}(\varepsilon) 
	{\rho}_S(\varepsilon){\rho}_S(\varepsilon+s\omega) 
	f_{0}(\varepsilon)[1-f_{0}(\varepsilon+s\omega)] \, . 
\end{equation}
and the reflection function  
\begin{align}
	\mathcal{R}^{s}_{\nu}(\varepsilon) 
	&=
	\mathcal{T}^{s}_{\nu}(\varepsilon) + 
		\textrm{Re}\left[ 
		\frac{\Delta}{\varepsilon}
		T_{\nu\nu}(\varepsilon)T^*_{\bar{\nu}\nu}(\varepsilon)  \right.  \nonumber \\
        	&+\!\! \left.  \frac{\Delta^2}{\varepsilon(\varepsilon+s\omega)}
		(T_{\nu\nu}(\varepsilon)T^*_{\bar{\nu}\bar{\nu}}(\varepsilon)+
		T_{\nu\bar{\nu}}(\varepsilon)T^*_{\bar{\nu}\nu}(\varepsilon)) \right]  	\, . 
\end{align}
The damping contribution associated to the quasiparticle reflections in the superconductor  $\gamma_{\mathrm{SS}}$ is independent of the applied voltage.
%
%
Moreover, for low temperatures compared to the superconducting gap $T \ll \Delta$, the population of quasiparticles 
above the gap is exponentially small such that we can disregard $\gamma_{\mathrm{SS}}$ for the rest of our discussion.

%
%
%
%
   
\section{Results for the phonon occupation}
\label{sec:occupation}

Applying a bias voltage, the electron current drives the oscillator to a nonequilibrium state  with phonon occupation
\begin{equation}
n = \langle b^\dagger b \rangle =
-\frac{1}{2}+\frac{i}{8\pi} \int d\varepsilon D^K(\varepsilon)
\simeq 
-\ \frac{\textrm{Im}\,\Pi^K(\omega)}{4\gamma_{}(\omega)}-\frac{1}{2} \, .
\end{equation}
In the limit $\gamma_{\textrm{tot}} \ll (\omega,\Gamma_N,\Gamma_S,T,eV)$ 
and separating the contribution of the intrinsic damping and the damping 
due to the charge-vibration interaction, the phonon occupation can be written as
\begin{align}
\bar{n} &=
\frac{ 
\gamma_{\mathrm{AR}} \, n_{\textrm{AR}} 
+
\gamma_{\textrm{NS}} \, n_{\textrm{NS}} 
+ (\gamma_{\mathrm{NR}} + \gamma_{\textrm{SS}} + \gamma_0) \, n_B(\omega)}
{ \gamma_{\mathrm{AR}}+ \gamma_{\textrm{NS}}  + \gamma_{\mathrm{NR}} + \gamma_{\textrm{SS}} + \gamma_0}
\label{eq:nbar}
\end{align}
with the Bose function $n_B(\omega)=[\mathrm{exp}(\omega/T)-1]^{-1}$.
The nonequilibrium occupation due to inelastic ARs reads [\onlinecite{Stadler:2016gl}] 
\begin{equation}
n_{\mathrm{AR}} = \frac{1}{\gamma_{\mathrm{AR}}} \sum_{\nu s } s \gamma_{\nu \bar{\nu}}^s  n_B(\omega\mathord+ \nu s\,2 e V) \, , 
\end{equation}
%
%
whereas the nonequilibrium occupation $n_{\textrm{NS}}$  is given by 
\begin{equation}
n_{\mathrm{NS}} = \frac{1}{\gamma_{\mathrm{NS}}}\sum_{\nu s } s \gamma_{\mathrm{NS}}^{s\nu}n_B(\omega+\nu s \, eV )
\end{equation}
with the Bose function shifted by $eV$ due to the tunneling of quasiparticles. 
Notice that the normal reflection (NR) and the quasiparticle reflections (SS) can drive the oscillator only to the 
thermal equilibrium.

For low temperature compared to the gap $T\ll\Delta$ and 
for the case $\gamma_0 \ll \gamma_{\mathrm{NR}}$, we can approximate 
$\gamma_{\mathrm{NR}} + \gamma_{\textrm{SS}} + \gamma_0 \simeq \gamma_{\mathrm{NR}}$.
Moreover,  in the high voltage limit $eV\gg T$ and for positive bias $eV>0$,
we have essentially that electrons from the normal lead are moving to the right superconducting lead 
(or equivalently the holes move from the right to the left).
In this case,  we can approximate the 
expression for $\bar{n}$ as
\begin{equation}
\bar{n} \simeq
\frac{
\gamma^{-}_{eh} \, + \,   \gamma^{-e}_{\textrm{NS}} \,  +  \,  \gamma_{\mathrm{NR}} n_B(\omega)}
{
\left( \gamma^{+}_{eh} + \gamma^{-}_{eh} \right)
 + 
 \left(  \gamma^{+e}_{\textrm{NS}} -  \gamma^{-e}_{\textrm{NS}} \right)
+ \gamma_{\mathrm{NR}} }
\label{eq:nbar_app}
\end{equation}
The Eq.~\eqref{eq:nbar_app}  generalizes the main result in Ref.~[\onlinecite{Stadler:2016gl}] as it takes into 
account the effects of the quasiparticles 
on the nonequilibrium phonon occupation of the mechanical oscillator.

\subsection{Single mode phonon occupation}
\label{sec:occupation_single_mode}  
\begin{figure}[t]			
    \begin{minipage}{\columnwidth}
  		\hspace{-5mm}\includegraphics[width=0.92\linewidth,angle=0.]{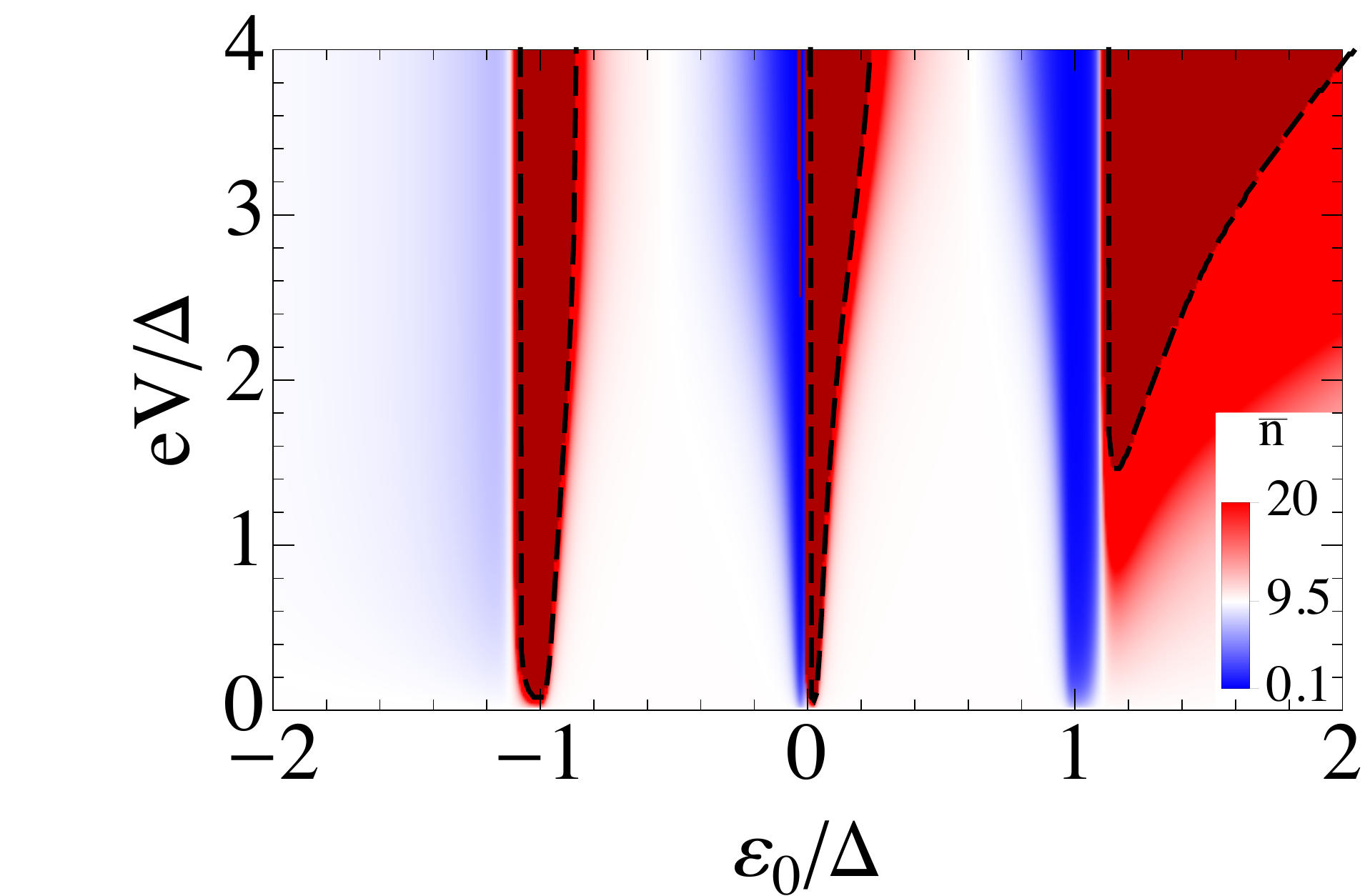} 
  	\end{minipage}
  	\caption{Phonon occupation as a function of bias voltage $eV$ and gate voltage $\varepsilon_0$ 
	at $T=10\omega$. The parameters are $\Gamma_N=\Gamma_S=0.2\omega$, $\lambda=0.1\omega$, 
	$\gamma_0=10^{-6}\omega$ and $\omega=0.05\Delta$. The dark red regions surrounded by dashed lines indicate a mechanical instability 	with $\gamma_{\mathrm{tot}}<0$.
  	}  
  	\label{fig:phonon_occupation}
\end{figure}

Figure \ref{fig:phonon_occupation} shows an example of the phonon occupation $\bar{n}$ as a function of  
$\varepsilon_0$ and source-drain voltage $eV$ for a single mode with frequency 
$\omega \ll \Delta$ and a temperature $T=10 \, \omega$. 
As anticipated in the introduction, we have: (i) regions of ground state cooling $\bar{n} \ll 1$, (ii) 
regions of strong energy pumping into the resonator $\bar{n} \gg 1$ and (iii) regions of a mechanical
instability in which the total electromechanical damping becomes negative $\gamma+\gamma_0 <0 $ and our 
perturbative method breaks down. 
In analogy to previous results in Ref.~[\onlinecite{Usmani:2007bu}], one expects that, going beyond the perturbative approach for the charge-vibration interaction and taking into account the anharmonicity of the resonator, such regions can host self-sustained oscillations.

%
%
%
%
%
%
%
%
%
%

The first striking observation in Fig.~\ref{fig:phonon_occupation}
is the weakly dependence  of the phonon occupation $\bar{n}$ on the bias-voltage whereas $\bar{n}$ is strongly dependent on $\varepsilon_0$.
In other words, the phonon occupation is determined by the bare dot's level $\varepsilon_0$, or the gate voltage, 
whereas no additional features appear at $eV \sim \Delta$. 
This behavior can be explained as we assume a sharp resonant regime with  a 
small coupling to the normal lead $\Gamma_N=0.1 \omega \ll \Delta$. 
As discussed in Sec.~\ref{subsec:Gdot}, the Andreev levels $E_A$ are formed within the superconducting gap 
whose position depends on $\varepsilon_0$ (and the coupling to the superconductor $\Gamma_S$) 
and whose small broadening is set by $\Gamma_N$. 
This translates into sharp peaks entering in the transmission functions in Eqs.~\eqref{eq:T_AR}, \eqref{eq:T_NR}, and \eqref{eq:T_s_nu}, 
and yield a saturation for the damping rates $\gamma_{\mathrm{AR}},\gamma_{\mathrm{NR}}$ and $ \gamma_{\textrm{NS}}$ 
as increasing the bias voltage.
Thus we divide the following discussion about the phonon occupation 
of Fig.~\ref{fig:phonon_occupation} into the deep subgap regime with   
$ \vert \varepsilon_0\vert \ll \Delta$, and a regime close to the gap $|\varepsilon_0| \sim \Delta$.
This scale separation is valid for $\omega \ll \Delta$.

In the first region at low energy, $|\varepsilon_0|\ll\Delta$, inelastic AR dominates [see Fig.~\ref{fig:damping_AR_NR}].
This subgap regime was discussed in Ref.~\onlinecite{Stadler:2016gl} in which we showed 
that resonances in both amplitudes of the transmission function in Eq.~\eqref{eq:T_AR} as well as interference 
between the two amplitudes result in: (i) an enhancement of the absorption rates compared to the emission rates for $\varepsilon_0 <0$ leading to  ground-state cooling, (ii) a large  phonon occupation eventually approaching a mechanical instability 
for $\varepsilon_0 > 0$. 
Here, we extend the previous result showing that 
$\bar{n}$ is controlled by the gate voltage even for bias voltages larger than 
the superconducting gap $eV > \Delta$. 
If additionally $|\varepsilon_0|\ll\Delta$, we can neglect the contribution of quasiparticles 
since the energy levels of the Andreev bound states lie deep inside the gap 
and quasiparticles inelastic transport occurs out of resonance.

\begin{figure}[t!]			
	\begin{minipage}{\columnwidth}
		\includegraphics[width=1\linewidth,angle=0.]{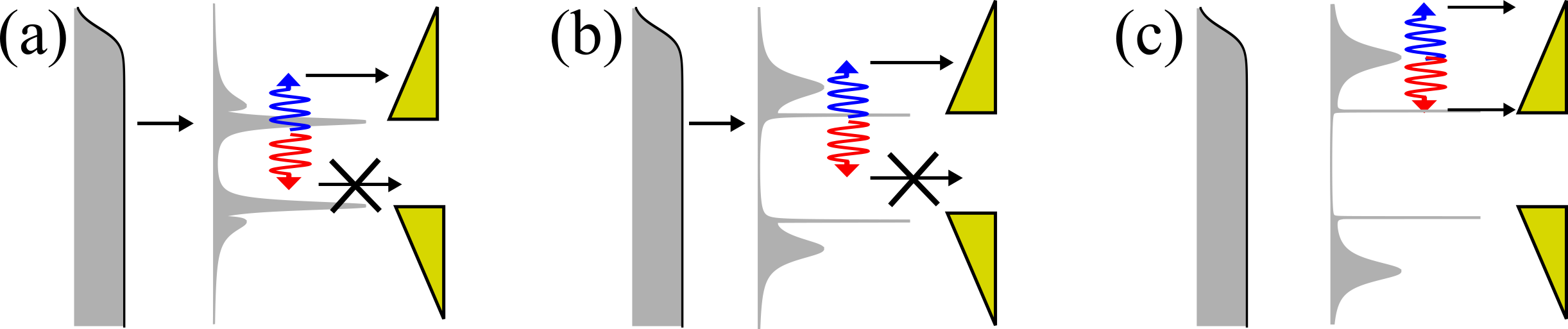} 
	\end{minipage}
	\caption{Schematic view of absorption and emission processes due to inelastic quasiparticle tunneling for the case $|eV-\Delta| \gg T$. 
		In (a) and (b) the emission of a vibrational energy quantum is essentially suppressed by the superconducting gap. 
		Figure (a) represents the case of $\varepsilon_0<\Delta$ and $E_A<\Delta$. 
		Figure (b) represents the case of $\varepsilon_0> \Delta$ but still  $E_A<\Delta$, in which the 
		density of states above the gap is enhanced due to the resonant level of the dot leading to a further 
		increase of the absorption processes.  
		In (c), for $\varepsilon_0+\omega \geq \Delta$, no Andreev bound states are formed in the dot and 
		the emission is enhanced due to the singular behavior of the superconducting density of state at the energy 
		$\varepsilon=\Delta$.
	}  
	\label{fig:sketch_DT}
\end{figure}

By contrast,  the inelastic tunneling of quasiparticles strongly controls the phonon occupation when the gate voltage approaches the superconducting gap $|\varepsilon_0| \simeq \Delta$ [see Fig.~\ref{fig:phonon_occupation}].
Indeed, in this region, inelastic quasiparticle tunneling are the leading processes as shown in Fig.~\ref{fig:damping}.
These processes, similarly to the inelastic ARs, yield a similar behavior of the nonequilibirium 
occupation with alternating regions of ground-state cooling, energy ``heating''  and instability.
The line separating the ground state cooling $\bar{n} \ll 1$ from the region $\bar{n} \gg 1$, 
is a narrow range in which the resonator is close to the thermal equilibrium. 
These results can be explained by considering the subtle interplay between the density of states 
$\rho_d(\varepsilon)$ of the quantum dot [see Fig.~\ref{fig:DensityOfStates}(a)], 
that acts as energy filter via the transmission coefficient $\mathcal{T}^{s}_{\nu}(\varepsilon)$ in Eq.~\eqref{eq:T_NR}, 
and  the behavior of the superconducting density of states with an energy gap and a divergence at $\varepsilon=\Delta$.
The basic idea is schematically reported in Fig.~\ref{fig:sketch_DT}.
Hereafter we discuss in detail the positive part $\varepsilon_0>0$ at $\varepsilon_0 \simeq \Delta$.
Similar arguments apply for the negative part  $\varepsilon_0<0$ at $\varepsilon_0 \simeq -\Delta$.

%
%
%
%
%
%
%
%
%
%

We start the discussion with the case  shown in Fig.~\ref{fig:sketch_DT}(a) when 
the Andreev states are well formed inside the dot and the energy level $\varepsilon_0 < \Delta$.
Tunneling from $N$ to $S$ with phonon absorption or phonon emission are possible for electrons with the appropriate
energy threshold, namely $\varepsilon \geq \Delta-\omega$ or $\varepsilon \geq \Delta+\omega$, respectively.
However, the density of states $\rho_d(\varepsilon)$  (and hence the transmission function $\mathcal{T}^{s}_{\nu}(\varepsilon)$)
is sharply peaked at the Andreev levels  such that one can assume that electron are essentially injected at energy  $\varepsilon=|E_A|$.  
In this case shown in Fig.~\ref{fig:sketch_DT}(a), the emission of a phonon is simply blocked due the superconducting gap.

Increasing $\varepsilon_0$, we approach the region $\varepsilon_0 > \Delta$. 
Here, beyond Andreev levels inside the gap, a broadened peak appears in $\rho_d(\varepsilon)$ 
 - and in the transmission  $\mathcal{T}^{s}_{\nu}(\varepsilon)$  - above the gap, 
reminiscent of the bare dot's level [see Fig.~\ref{fig:DensityOfStates}(a) and Fig.~\ref{fig:sketch_DT}(b)].
In this case, beyond the fact that phonon emission is still  blocked due the superconducting gap, 
the absorption rate is in addition enhanced due to the increase in $\rho_d(\varepsilon)$ and $\mathcal{T}^{s}_{\nu}(\varepsilon)$
 at $\varepsilon >\Delta$.

Finally, further increasing $\varepsilon_0$, the peak of the functions ${\rho}_S(\varepsilon)$ or $\mathcal{T}^{s}_{\nu}(\varepsilon)$ moves 
far away from the gap edge such that the condition $\varepsilon_0=\Delta+\omega$ is reached.
In this case, electrons from the normal lead N are again filtered by the dot at energy  $\varepsilon=\varepsilon_0$ 
and phonon emission are greatly enhanced in this case due to the divergence of ${\rho}_S(\varepsilon)$ 
at the gap edge, viz. $\gamma^-_{\textrm{NS}} > \gamma^{+}_{\textrm{NS}}$, leading to the instability.

%
%
\subsection{Multimode phonon ground-state cooling}
\label{sec:occupation_multi_mode}
In this section, we illustrate that several nondegenerate mechanical modes can be cooled to the ground state 
in the N-QS-S system.

We generalized the Hamiltonian Eq.~(\ref{eq:H}) considering several vibrational eigenmodes of the resonator.
The frequencies of these vibrational modes are denoted by $\omega_k$ and the bosonic annihilation and creation operators for each mode
by $\hat{b}_k^{\phantom{g}}$ and $\hat{b}_k^{\dagger}$. 
The Hamiltonian reads: 
\begin{equation}
\hat{H} \!=\! 
\hat{H}_N \!+\! \hat{H}_S \!+\! \hat{H}_t \!+\! \varepsilon_0 \hat{n}_d  
\!+\! \sum_k \left[ \lambda_k  \hat{n}_d (\hat{b}^\dagger_k+\hat{b}^{\phantom{\dagger}}_k )
+ \omega_k \hat{b}^\dagger_k\hat{b}^{\phantom{\dagger}}_k\right] \, .
\label{eq:H_multi}
\end{equation}
To illustrate the multimode cooling, we  assume  a linear low-frequency spectrum by $\omega_k=k\omega$ with $k=1,2,\dots,N$ although  this assumption is not essential.
The calculation of the multimode phonon occupation can be limited by considering a finite number of modes, as we have a natural cut-off given by the temperature. 
For the case $T=10\omega$ we choose $N=7$  modes.
High-frequency modes with $\omega \gtrsim T$ are close to the ground-state. 
The nonequilibrium value $\bar{n}_k$ for each mode is calculated by Eq.~\eqref{eq:nbar} from which 
we can plot the total mechanical energy defined by $E_{\mathrm{tot}}=\sum_{k=1}^{N} \omega_k \bar{n}_k$. 
Such a approach holds since, in the regime of weak charge-vibration coupling, 
we can neglect the intercorrelation effect between the different modes.
%

\begin{figure}[t!]			
	\begin{minipage}{\columnwidth}
		\includegraphics[width=0.92\linewidth,angle=0.]{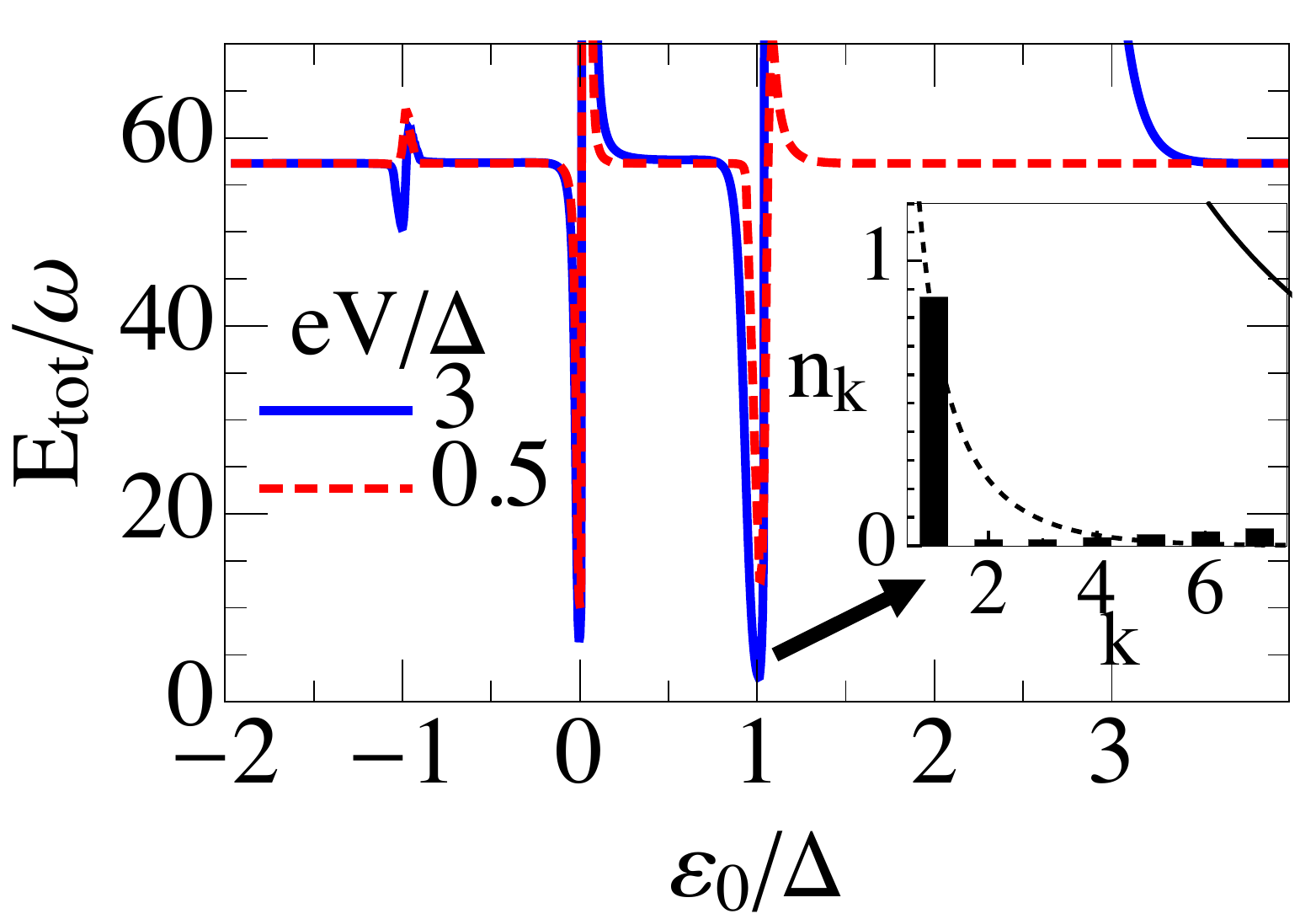} 
	\end{minipage}
	\caption{Multimode phonon occupation. Total mechanical energy  
		$E_{\mathrm{tot}}=\sum_k \hbar \omega_k \bar{n}_k$ for $k=7$ modes as a function gate voltage $\varepsilon_0$ at $T=10\omega$ 
		for two different bias voltages. 
		The parameters are $\Gamma_N=\Gamma_S=0.2\omega$, $\omega=0.01\Delta$, $\gamma_0=10^{-6}\omega$ 
		The total mechanical energy at equilibrium is given by $E_{\mathrm{tot}}\approx57.2\omega$. The dashed line in the inset corresponds to the Bose distribution $n_B(\omega_k)$ at $T=1.2\omega$. 
	}  
	\label{fig:multimode_occupation}
\end{figure}

In Fig.~\ref{fig:multimode_occupation} we show the multimode phonon occupation at bias voltage larger than the gap $eV > \Delta$ and 
as a function of $\varepsilon_0$.
Similarly to the discussion in Sec.~\ref{sec:occupation_single_mode} of the phonon occupation for a single mode, we found that the total energy  $E_{\mathrm{tot}}$  
has a weak bias-voltage dependence and its behavior is controlled by the gate voltage $\varepsilon_0$. 

As can be seen from Fig.~\ref{fig:multimode_occupation}, the thermal equilibrium value of the 
total mechanical energy $E_{\mathrm{tot}}\approx57.2\omega$ is 
approached if the gate voltage is much larger than the gap, $\vert \varepsilon_0 \vert \gg \Delta$.
Figure~\ref{fig:multimode_occupation} has a similar behavior as the single mode results $\bar{n}$ with features in the deep subgap region $|\varepsilon_0| \ll \Delta$
and close to the gap edges $|\varepsilon_0| \sim\Delta$.
Around these ranges, the total energy shows dips below the thermal equilibrium value followed by sharp increases of the multimode phonon occupation.

When the gate voltage is close to $\varepsilon_0 \simeq -\Delta$ and for $eV=3\Delta$,
the total energy has a small dip (weak cooling) followed by a increase due to inelastic tunneling of quasiparticles. 
In this case, since the temperature is small compared to the superconductor gap $T \ll \Delta$, 
the states in the superconductor are almost filled and the phonon occupations of the several modes are only slightly reduced.

As discussed in a previous work in Ref.~[\onlinecite{Stadler:2016gl}], 
for negative $\varepsilon_0<0$ and  $|\varepsilon_0| \ll \Delta$ , 
all the vibrational modes are strongly cooled 
by inelastic-vibration assisted Andreev reflections. 
Due to an interference mechanism, such a cooling is possible without matching a resonant condition between levels of the quantum dot 
and the vibrational frequency. 
Here we demonstrate that this mechanism is working even at  bias voltage larger than the gap.
%

Interestingly, the  phonon occupation of several modes is strongly reduced even 
closely below the superconducting gap,  for $\varepsilon_0\simeq\Delta$, 
where the dominant processes are the inelastic tunneling of quasiparticles,  
similarly as discuss in the previous section. 

%
%
%
%
%
%
%
%
%

%
%
%
%
\section{Results for the current}
\label{sec:current}

In this section, we calculate the current to the leading order in the charge-vibration coupling for a single mode with frequency $\omega$.

Using the spinor notation in Nambu space in the normal lead
$\Psi_{k}^\dagger = ( c_{k\uparrow}^\dagger \, c_{-k\downarrow})$ 
and the number operator
$
\hat{N} = \frac{1}{2}\sum_{k} \Psi_{k}^\dagger \hat{\tau}_z \Psi_{k}
$,
with the Pauli matrix $\hat{\tau}_z$, 
the current through the normal lead $I  = -e \langle \mathrm{d}\hat{N}_n/\mathrm{d}t \rangle $  
can be expressed as 
\begin{equation}
I = 
\frac{e}{2\hbar}\int \frac{d\varepsilon}{2\pi} \,\mbox{Re}\,\mbox{Tr} [\hat{\tau}_z({\hat{\mathcal{G}}}^K(\varepsilon) \hat{\Sigma}_{N}^A+{\hat{\mathcal{G}}}^R(\varepsilon) \hat{\Sigma}_{N}^K(\varepsilon) ) ] \, ,
\label{eq:Il}
\end{equation}
in which $\hat{\mathcal{G}}^K(\varepsilon)$ and $\hat{\mathcal{G}}^R(\varepsilon)$ refer to the dot Green's functions 
and they are defined similarly as in Sec.~\ref{subsec:Gdot}.
They correspond to the exact Green's function including the coupling to the leads and the charge-vibration interaction. 
To calculate the current in Eq.~\eqref{eq:Il} we perform a perturbation expansion of these Green's functions to the leading order in 
the charge-vibration coupling $\lambda$ [\onlinecite{Egger:2008hha},\onlinecite{Stadler:2015dga}]. 
Hence the current can be decomposed into a elastic current, 
an elastic correction and an inelastic term as 
\begin{equation}
I = I_0+I_{\mathrm{ec}}^{(\lambda^2)}+I_{\mathrm{in}}^{(\lambda^2)}
\end{equation}
with
\begin{align}
I_0 &\mathord= 
e \int \frac{d\varepsilon}{2\pi} \,\mbox{Re}\,\mbox{Tr} [\hat{\tau}_z (\hat{G}^K \hat{\Sigma}_N^A\mathord+\hat{G}^R\hat{\Sigma}_N^K) ] 
\label{eq:I0}\\
I_{\mathrm{ec}}^{(\lambda^2)} &\mathord= 
e \int \frac{d\varepsilon}{2\pi} \, 
\mbox{Re}\,\mbox{Tr} [\hat{\tau}_z (\hat{G}^R \hat{\Sigma}_{\mathrm{vib}}^R \hat{G}^K\Sigma_N^A
\mathord+\hat{G}^K\hat{\Sigma}_{\mathrm{vib}}^A\hat{G}^A\hat{\Sigma}_N^A \nonumber  \\ 
&\mathord+\hat{G}^R \hat{\Sigma}_{\mathrm{vib}}^R \hat{G}^R (\hat{\Sigma}_N^R
\mathord-\hat{\Sigma}_N^A\mathord+\hat{\Sigma}_S^R\mathord-\hat{\Sigma}_S^A)\hat{G}^A\hat{\Sigma}_N^K)] \\
I_{\mathrm{in}}^{(\lambda^2)} &\mathord= e \int\frac{d\varepsilon}{2\pi} \,
\mbox{Re}\,\mbox{Tr}[\hat{\tau}_z(\hat{G}^R \hat{\Sigma}_{\mathrm{vib}}^K \hat{G}^A \hat{\Sigma}_N^A
\mathord+\hat{G}^R \hat{\Sigma}_{\mathrm{vib}}^R\hat{G}^A \hat{\Sigma}_N^K ) ]  \label{eq:Iin}
\end{align}
and the self-energies associated to the charge-vibration interaction $(\sim \lambda^2)$
\begin{align}
\hat{\Sigma}_{\mathrm{vib}}^{R/A}(\varepsilon)
&=
i \frac{\lambda^2}{2} \int \frac{d\varepsilon^\prime}{2\pi} D^{R/A}(\varepsilon^\prime) \hat{\tau}_z \hat{G}^K(\varepsilon-\varepsilon^\prime)\hat{\tau}_z  \nonumber \\
& \hspace{2cm}-
 \hat{\tau}_z\mathrm{Tr}\,[\hat{\tau}_z \hat{G}^K(\varepsilon^\prime)] D^{R,A}(0) \, , \\
\hat{\Sigma}_{\mathrm{vib}}^{K}(\varepsilon)
&= i \frac{\lambda^2}{2} \int \frac{d\varepsilon^\prime}{2\pi} \sum_{\xi=R,A,K } D^{\xi}(\varepsilon^\prime)  \hat{\tau}_z\hat{G}^{\xi}(\varepsilon-\varepsilon^\prime)  \hat{\tau}_z \, .
\end{align}
The elastic current $I_0$ shows a peak at $\varepsilon_0=0$ as a function of the gate voltage which is associated to the Andreev reflections. 
%
%
In the deep subgap regime, this current can be simplified to
\begin{equation}
I_0= 8e \Gamma_N^2 \int d\varepsilon \,   \vert F(\varepsilon) \vert^2  [f_+(\varepsilon)-f_-(\varepsilon)] \,.
\end{equation}
%
%
In addition, $I_0$ increases close to the gap edges at $\varepsilon_0\simeq \vert \Delta \vert$ of the gate voltage, 
where quasiparticle tunneling occurs and contributes to transport. 

The term $I_{\mathrm{ec}}^{(\lambda^2)}$ represents the corrections to the elastic part and shows no additional features with respect to $I_0$.
On the other hand, the inelastic term of the current $I_{\mathrm{in}}^{(\lambda^2)}$ has different features.
This term can be approximated in the high-voltage limit as 
\begin{equation}
\label{eq:I_in_app}
I_{\mathrm{in}}^{(\lambda^2)} / e \simeq
\left( \bar{n} + 1 \right) \left( \gamma_{\mathrm{AR}}^{-}  + \gamma_{\textrm{NS}}^{-}   \right)
+
\bar{n} \left( \gamma_{\mathrm{AR}}^{+}  +  \gamma_{\textrm{NS}}^{+}   \right)
\end{equation}
for $eV \gg T$. 
The inelastic current is directly related to the average phonon occupation.
In the following, we discuss two different cases: First we consider the vibrational states at thermal equilibrium 
$\bar{n}\simeq n_B(\omega) $ and second the case of vibrational states out of equilibrium such that $\bar{n}$ is completely controlled by the charge passing through the dot.

%
%
%
%
\begin{figure}[t]			
	\begin{minipage}{\columnwidth}
		\includegraphics[width=0.8\linewidth,angle=0.]{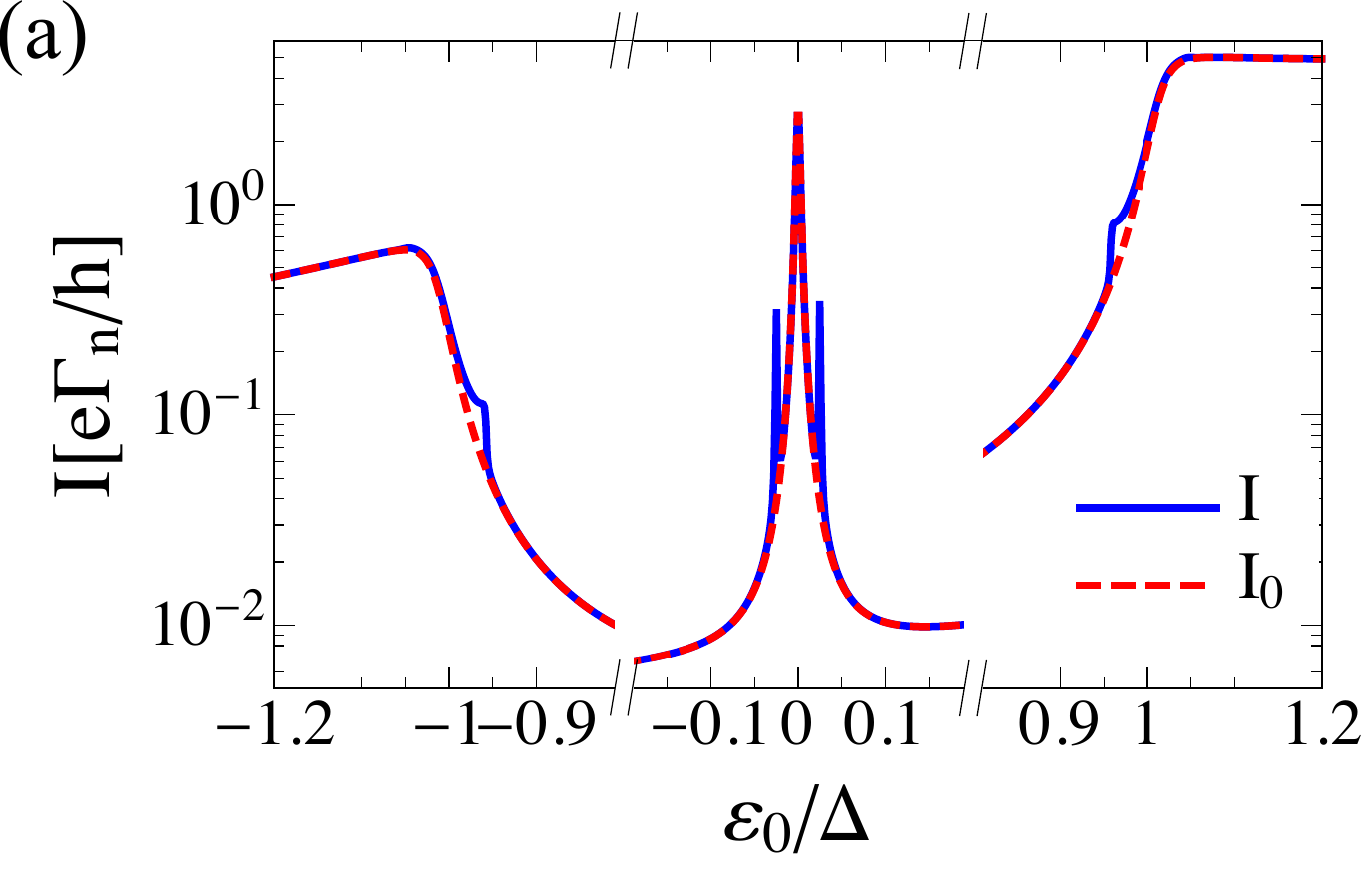} 
	\end{minipage}
	\begin{minipage}{\columnwidth}
		\includegraphics[width=0.8\linewidth,angle=0.]{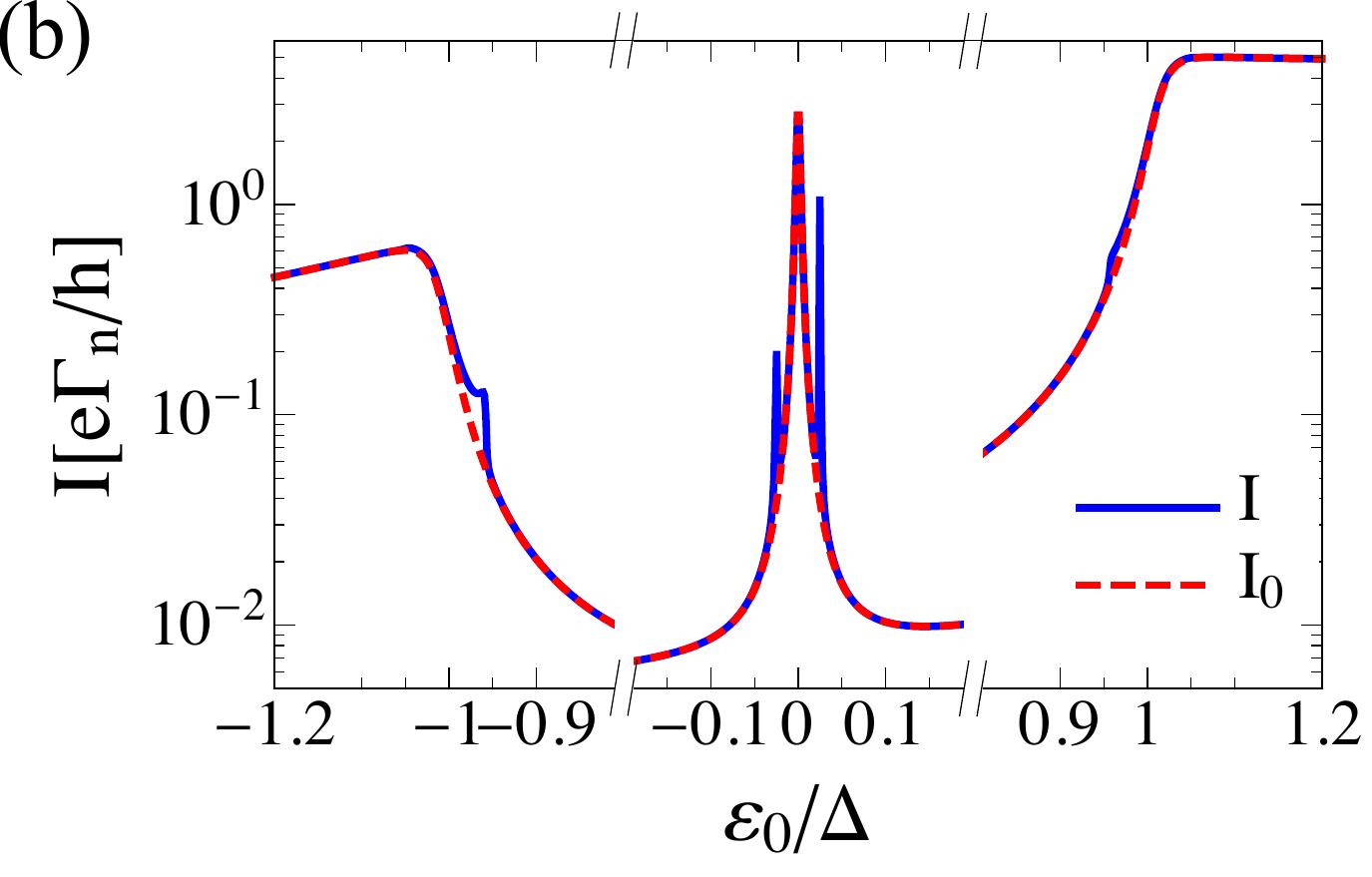} 
	\end{minipage}
	\caption{Total current $I$ and elastic current $I_0$ as a function of $\varepsilon_0$ at $\omega=0.05\Delta$. In (a), the resonator is in thermal equilibrium with large intrinsic damping such that $\bar{n}=n_B(\omega)$. In (b), the resonator is in a nonequilibrium state with $\lambda=0.015\omega$ and $\gamma_0=10^{-4}\omega$. The other parameters are $\Gamma_N=0.015\omega$, $\Gamma_S=0.05\omega$, $T=10\omega$ and $eV=10\Delta$. 
	}  
	\label{fig:current}
\end{figure}
%
%
%
%
%

As example of this comparison, we show in Fig.~\ref{fig:current}  
the current as a function of $\varepsilon_0$ at large bias voltage $eV=10\Delta$ 
for a resonator at equilibrium (a) and in a nonequilibrium state (b).

In both cases, we observe that the inelastic current shows peaks which can be divided into contribution due to inelastic Andreev reflection and quasiparticle tunneling. 
Close to the zero-bias peak $\varepsilon_0=0$, 
inelastic-vibration assisted Andreev reflection at $\varepsilon_0=\pm \omega/2$ give rise to transport: 
The peak at $\varepsilon_0=-\omega/2$ is attributed to the absorption of a phonon 
whereas the peak at $\varepsilon_0=\omega/2$ to the emission of one phonon [\onlinecite{Stadler:2016gl}].

Additionally to these peaks, we also found peaks close to the gap edges where 
inelastic quasiparticles tunneling give a contribution to transport 
at $E_A=\pm \vert \Delta  \pm \omega\vert$,  using the same argument 
discussed at the end of  the Sec.~\ref{sec:occupation_single_mode} [see Fig.~\ref{fig:sketch_DT}].
The peaks at $E_A= \vert\Delta -\omega\vert$ corresponds to the absorption of one phonon and 
$E_A= - \vert\Delta -\omega\vert$ to the emission of one phonon, namely 
the role of the negative and positive peaks are inverted compared to peaks associated to the ARs.

The difference behaviors of the peaks  in  Fig.~\ref{fig:current}(a) and  Fig.~\ref{fig:current}(b) 
can be explained by the phonon occupation Fig.\ref{fig:phonon_occupation} 
with the help of the expression of the inelastic current in Eq.~\eqref{eq:I_in_app}.

When the vibration is at thermal equilibrium  as in Fig.~\ref{fig:current}(a), 
both kind of the peaks due to the inelastic processes involving ARs and quasi-particles are almost of the same height. 
The peak due emission are slightly larger than the peaks due to emission of one phonon due factor $n_B(\omega)+1$ in front of emission processes compared to the factor $n_B(\omega)$ in front of the absorption processes.

By contrast, in Fig.~\ref{fig:current}(b) the resonator is in a nonequilibrium state and the paired sidepeaks 
are of different height: 
the peaks which are associated to the phonon absorption are suppressed compared to the peaks associated 
to the phonon emission. 
This result implies that the heights of the inelastic peaks lying in two different range of the gate voltage 
are correlated and they can be used to check if the resonator is in a nonequilibrium state.
Using the same argument discussed in Ref.~[\onlinecite{Stadler:2016gl}], 
the ratio between the brightness  of paired peaks  - defined as the underlying area for each peak -
can be used to extract information about the phonon occupation of the non equilibrated resonator.

%
%
%
%
\section{Conclusion}
\label{sec:conclusions}
We studied the phonon occupation and the quantum transport of a quantum dot embedded between a normal-conducting and a superconducting lead showing that the vibrational states of the resonator are controlled by applying a bias-voltage. The enhancement or suppression of the phonon occupation depends strongly on the resonant level of the quantum dot. When the energy of the resonant level is well inside the superconducting gap, inelastic vibration-assisted Andreev reflections drive the resonator to a nonequilibrium state. At gate-voltages close to the superconducting gap edge, inelastic tunneling of quasiparticles control the state of the resonators. As a result we obtained that inelastic vibration-assisted Andreev reflections and inelastic tunneling of quasiparticles cool the vibrational state of the resonator to the ground-state even for several vibrational modes. The current shows characteristic signatures of the inelastic tunneling processes and can be exploited to detect the resonator in a nonequilibrium state. Similar inelastic processes were observed in recent experiments [\onlinecite{Schindele:2014fm,Gramich:2015dk,Gramich:2016bs,Gramich:2016arxiv}] suggesting that our proposal in within the reach of current research.

\acknowledgments 
We acknowledge J.~C. Cuevas and E. Scheer for interesting discussions and useful comments.
This research was supported by the Zukunftskolleg of the University of Konstanz and 
by the DFG through the collaborative research center SFB 767.
%

\bibliography{references}


\end{document}